\documentclass[twocolumn,epjc3]{svjour3mod}

\usepackage{amsmath,amssymb,amsfonts,dcolumn,mathcomp,slashed,dsfont}
\usepackage{graphicx}
\usepackage{booktabs}
\usepackage{bm}
\usepackage{xcolor}
\usepackage{csquotes}
\usepackage{units}
\usepackage{flushend}
\usepackage{cite}
\usepackage[colorlinks,citecolor=blue,urlcolor=blue,linkcolor=blue]{hyperref}
\usepackage{ragged2e}

\hypersetup{
    pdfauthor={Simon Mutke, Leon A. Heuser, Ingrid Dax, Bastian Kubis, Stefan Leupold}, 
    pdftitle={Dispersive Analysis of D- and B-Meson Form Factors with Chiral and Heavy-Quark Constraints}
}

\usepackage{braket}
\usepackage{tikz}
\usepackage{enumitem}
\usepackage{subcaption}

\apptocmd\thebibliography\justifying{}{}

\newcommand{\BR}{\mathcal{B}}
\newcommand{\Fpi}{F_\pi}
\newcommand{\Order}{\mathcal{O}}
\newcommand{\tr}{\text{Tr}}
\newcommand{\disc}{{\rm disc}\,}
\newcommand{\disca}{{\rm disc}_{\rm an}\,}
\newcommand{\dd}{{\rm d}}
\newcommand{\ii}{{\rm i}}
\newcommand{\mpi}{M_{\pi}}
\newcommand{\mV}{M_V}
\newcommand{\mP}{M_P}
\newcommand{\zV}{z_V}
\newcommand{\zP}{z_P}

\renewcommand{\Im}{{\rm Im}\,}
\newcommand{\ie}{\textit{i.e.}}

\newcommand{\cf}{\textit{cf.}}
\newcommand{\pcm}{p_{\pi}}
\newcommand{\pz}{p_{M}}
\newcommand{\abs}[1]{\left|#1\right|}

\newcommand{\grhopipi}{\ensuremath{g_{\rho\pi\pi}}}
\newcommand{\grhomm}{\ensuremath{g_{\rho MM}}}
\newcommand{\grhomstarm}{\ensuremath{g_{\rho M^* M}}}
\newcommand{\grhomstarmstar}{\ensuremath{g_{\rho M^* M^*}^{(1)}}}
\newcommand{\frhomstarmstar}{\ensuremath{g_{\rho M^* M^*}^{(2)}}}
\newcommand{\hrhomstarmstar}{\ensuremath{g_{\rho M^* M^*}^{(3)}}}
\newcommand{\hrhodstardstar}{\ensuremath{g_{\rho D^* D^*}^{(3)}}}
\newcommand{\hrhobstarbstar}{\ensuremath{g_{\rho B^* B^*}^{(3)}}}

\newcommand{\GeV}{\,\text{GeV}}
\newcommand{\MeV}{\,\text{MeV}}
\newcommand{\keV}{\,\text{keV}}
\newcommand{\bsp}{\begin{sloppypar}}
\newcommand{\esp}{\end{sloppypar}}

\def\XXint#1#2#3{{\setbox0=\hbox{$#1{#2#3}{\int}$}
     \vcenter{\hbox{$#2#3$}}\kern-0.5\wd0}}

\renewcommand{\emph}[1]{\textit{#1}}
\renewcommand{\vec}[1]{\boldsymbol{#1}}

\def\vecsign{\mathchar"017E}
\def\dvecsign{\smash{\stackon[-2.17pt]{\vecsign}{\rotatebox{180}{$\vecsign$}}}}
\def\dvec#1{\def\useanchorwidth{T}\stackon[-4.2pt]{#1}{\,\dvecsign}}
\usepackage{stackengine}
\stackMath

\begin{document}

\title{\boldmath Dispersive Analysis of $D$- and $B$-Meson Form Factors with Chiral and Heavy-Quark Constraints}

\author{Simon Mutke\thanksref{addr1,e1}
\and
Leon A.\ Heuser\thanksref{addr1,e2}
\and
Ingrid Dax\thanksref{addr1}
\and
Bastian Kubis\thanksref{addr1} 
\and
\hbox{Stefan Leupold}\thanksref{addr2}}

\thankstext{e1}{e-mail: mutke@hiskp.uni-bonn.de}
\thankstext{e2}{e-mail: heuser@hiskp.uni-bonn.de}

\institute{Helmholtz-Institut f\"ur Strahlen- und Kernphysik (Theorie) and
   Bethe Center for Theoretical Physics,
   Universit\"at Bonn, 
   53115 Bonn, Germany\label{addr1}
\and
   Institutionen f\"or fysik och astronomi, 
   Uppsala universitet,
   Box 516, 75120 Uppsala, Sweden\label{addr2}
}

\date{}

\maketitle

\newpage
\newpage

\begin{abstract}
We analyze the isovector vector form factors of $D$, $D^*$, $B$, and $B^*$ mesons at low energies.  We employ all constraints due to chiral and heavy-quark symmetry, and include the physics of resonant pion--pion rescattering in a model-independent way, using dispersion theory.  Special attention is paid to the analytic properties of these form factors, which include anomalous thresholds due to triangle diagrams that are located on the physical Riemann sheets in some of the form factors.  We extract the couplings of the $\rho(770)$ resonance to all these heavy mesons by determining the appropriate pole residues.
\end{abstract}

\section{Introduction}

Quantum chromodynamics (QCD) describes the strong interaction as a multi-scale phenomenon. The dynamical scale, $\Lambda_{\rm QCD}$, is set by the momentum exchange for which the QCD running coupling constant changes from large to small. Only for a small coupling can quark--gluon perturbation theory be used. The masses of the up and down quarks are much lighter than the dynamical QCD scale; on the other hand, the masses of the charm and bottom quarks are much heavier\footnote{In this work, we do not consider top quarks explicitly.} than $\Lambda_{\rm QCD}$. This scale separation suggests the use of effective field theories (EFTs) to deal with the strong interaction in the non-perturbative regime. The pertinent symmetries are of central importance for these EFTs. 

In the chiral limit, \ie,\ neglecting the up and down quark masses, QCD has a chiral symmetry that is spontaneously broken. This gives rise to isovector pseudoscalar Goldstone bosons, the pions. In the heavy-quark limit, \ie,\ neglecting the inverse of the mass of a charm or bottom quark, one can identify a spin--flavor symmetry. Both symmetries play a role for the structure and dynamics of $B$ and $D$ mesons. A systematic treatment of the consequences of the chiral symmetry and of corrections in powers of the light quark masses is achieved by chiral perturbation theory~\cite{Weinberg:1978kz,Gasser:1983yg,Wise:1992hn,Yan:1992gz,Burdman:1992gh,Casalbuoni:1996pg}. A systematic treatment of the spin--flavor symmetry and of corrections in powers of the Compton wavelength of heavy quarks is provided by heavy-quark effective theory~\cite{Isgur:1989vq,Isgur:1990yhj,Grinstein:1990mj,Eichten:1989zv,Georgi:1990um,Manohar:2000dt}. In particular, the approximately realized symmetries suggest that the heavy quarks can be treated non-relativistically with the feature that spin-independent interactions are more important than spin interactions; they also suggest that the physics associated with the light quarks is highly relativistic with the usual features of particle production and annihilation. If one explores the structure of $D$ and $B$ mesons on a length scale that corresponds to the energies sufficient for particle production, one has to deal with relativistic (and non-perturbative) many-body aspects. 

The minimal quark content of $D$ mesons (anti-$B$ mesons) is a charm (bottom) quark and a light anti-quark. Albeit defining important aspects of these heavy--light mesons, this does not provide a complete picture of their structure. At large distances, the relativistic physics of the (approximately massless) pions influences the structure of the heavy--light mesons. The purpose of this work is to explore this long-range-structure aspect. We focus on the isovector electromagnetic form factors of $B$ and $D$ mesons. Consistently with the heavy-quark symmetry, we treat pseudoscalar and vector heavy--light mesons on equal footing. Consequently, we address the charge form factor of the pseudoscalar $B$ meson ($D$ meson), the three electromagnetic form factors of the vector $B^*$ meson ($D^*$ meson), and the transition form factor of the pseudoscalar to the vector meson. 

Let us rephrase the important aspects in terms of relevant hadronic degrees of freedom. 1.\ Heavy-quark symmetry demands to include both pseudoscalar and vector heavy--light mesons. 2.\ Spontaneously broken chiral symmetry brings in the pions as the agents that can transport information over large distances and therefore shape the long-range structure of the heavy--light mesons. 3.\ Finally, as expressed in the vector-meson-dominance picture~\cite{Sakurai:1960ju,Sakurai:1969fba,Landsberg:1985gaz,Meissner:1987ge,Klingl:1996by,Fang:2021wes}, electromagnetic properties of strongly interacting objects are largely influenced by those hadrons that have the same quantum numbers as photons, the neutral vector mesons. With our focus on isovector form factors, the $\rho$ meson is singled out as an important player.\footnote{Throughout this paper, we will use $\rho \equiv \rho(770)$, $\rho^\prime \equiv \rho(1450)$, and $\rho^{\prime\prime} \equiv \rho(1700)$ as abbreviations.}
 
The first two aspects are covered by heavy-quark chiral perturbation theory~\cite{Wise:1992hn,Yan:1992gz,Burdman:1992gh,Casalbuoni:1996pg}, but the physics of $\rho$ mesons is not. On the other hand, the main source of information about the $\rho$ meson comes from the pion--pion $P$-wave phase shift~\cite{Ananthanarayan:2000ht,Garcia-Martin:2011iqs} and the pion vector form factor~\cite{Colangelo:2018mtw}. In the spirit of Refs.~\cite{Kang:2013jaa,Granados:2017cib,Alarcon:2017asr,Leupold:2017ngs,Junker:2019vvy,Lin:2022dyu,Alvarado:2023loi,Aung:2024qmf,An:2024pip}, we use dispersion theory to relate the form factors of the heavy--light mesons to the pion vector form factor and the reaction amplitudes for the scattering of pions off heavy--light mesons in a model-independent way. The rescattering of pions in their $P$~wave is again taken care of by dispersion theory, while the ingredient of the ultimate coupling between pions and heavy--light mesons is provided by heavy-quark chiral perturbation theory at next-to-leading order in the chiral counting and leading order in the heavy-quark expansion. 

Beyond the direct aim to obtain a deeper insight into the long-range-structure aspects of heavy--light mesons, we provide the grounds for a more detailed exploration of the structure of composite objects beyond the single-hadron level. Several of the recently discovered $X$, $Y$, $Z$ states with hidden charm or bottomness~\cite{Olsen:2017bmm,Brambilla:2019esw} lie close in mass to the thresholds of pairs of heavy--light mesons. Therefore, these two-particle states strongly influence or even dominate the structure of such $X$, $Y$, or $Z$ states~\cite{Guo:2017jvc,Liu:2019zoy,Chen:2022asf}. To understand the structure of such composite objects with hidden charm or bottomness, it is helpful to know about the structure of the underlying building blocks, the heavy--light meson states with open charm or bottomness. Actually, two distinct aspects come into play here. First, a possible future investigation of the electromagnetic form factors of $X$, $Y$, or $Z$ states can profit from the knowledge about the corresponding form factors of the heavy--light mesons, which serve as building blocks. Second, on a more quantitative level, the binding forces for the mere formation of $X$, $Y$, or $Z$ states can be classified according to their respective interaction range. The longest-range force is clearly provided by the one-pion exchange, while (correlated) two-pion exchanges are the next in line~\cite{Guo:2017jvc,Meng:2022ozq,Chacko:2024cax}. Here, two pions correlated to $\rho$ mesons play an important role~\cite{Voloshin:1976ap}. Our framework covers exactly such interactions. As we demonstrate below, this can be translated to three-point coupling constants for heavy--light mesons coupled to $\rho$ mesons, rigorously defined via residues on the second Riemann sheet.  In the present work, however, we focus on the electromagnetic form factors of heavy--light mesons and leave applications to $X$, $Y$, $Z$ states for the future. 

This manuscript is organized as follows.  In Sect.~\ref{sec:HMFFs}, we give all necessary definitions for heavy-meson form factors coupling to an isovector vector current.  Section~\ref{sec:scattering} describes the construction of the amplitude for heavy-meson--antimeson annihilation into a pair of pions, based on heavy-meson chiral perturbation theory and subsequent dispersive unitarization.  In Sect.~\ref{sec:formfactors}, the resulting form factor dispersion relations are discussed and evaluated in detail.  The extraction of the various coupling constants of the $\rho$ meson is demonstrated in Sect.~\ref{sec:rho-couplings}, before we summarize in Sect.~\ref{sec:summary}.  Some technical details are relegated to the appendices.

\section{Heavy-Meson Form Factors}\label{sec:HMFFs}
Our goal is to obtain a dispersive framework for heavy-meson form factors describing the transition $M^{(*)} \bar{M}^{(*)} \to \gamma^*$ of either pseudoscalar mesons $M$ or vector mesons $M^{*}$ into a virtual photon,
\begin{align}
    \epsilon_\mu^*(\vec{q},\lambda) \bra{0} j^\mu(0) \ket{M^{(*)}(p_1)\,\bar{M}^{(*)}(p_2)}\,,
\end{align}
where $\epsilon_\mu(\vec{q},\lambda)$ is the polarization vector of the photon and $j^\mu(0)=\sum_{q} Q_q \bar{q} \gamma^\mu q$ is the electromagnetic current. We focus on a low-energy description for both $D$-meson and $B$-meson form factors, taking into account pion--pion ($\pi\pi$) intermediate states. This leads to the isovector parts of the form factors, which are expected to be dominated by the $\rho$ resonance. The isospin decomposition of these form factors is given by
\begin{align}
\begin{alignedat}{2}
    M^{(*)+}\,M^{(*)-} &\to \gamma^*:&\quad F^+ &= F^\text{IS} + F^\text{IV}\,, \\
    M^{(*)0}\,\bar{M}^{(*)0} &\to \gamma^*:&\quad F^0 &= F^\text{IS} - F^\text{IV}\,,
\end{alignedat}
\end{align}
which means that the isoscalar (IS) and isovector (IV) parts are obtained by
\begin{align} \label{eq:isospin_decomposition}
    F^\text{IS} &= \frac{1}{2} \bigl( F^+ + F^0 \bigr)\,, & 
    F^\text{IV} &= \frac{1}{2} \bigl( F^+ - F^0 \bigr)\,.
\end{align}
Throughout the paper we work in the isospin limit (except when stated otherwise) with the following (averaged) masses~\cite{ParticleDataGroup:2024cfk}:
\begin{align}
    M_{D} &\equiv \frac{M_{D^+}+M_{D^0}}{2} = 1.86725(4)\GeV\,, \notag\\
    M_{D^*} &\equiv \frac{M_{D^{*+}}+M_{D^{*0}}}{2} = 2.00856(4)\GeV\,, \notag\\
    M_{B} &\equiv \frac{M_{B^+}+M_{B^0}}{2} = 5.27956(5)\GeV\,, \notag\\
    M_{B^*} &\equiv \frac{M_{B^{*+}}+M_{B^{*0}}}{2} = 5.32475(20)\GeV\,, \notag\\
    \mpi &\equiv M_{\pi^+} = 139.57039(18)\MeV\,.
\end{align}
Since only the charged pions contribute to the isovector combination, we do not average the pion mass.

\subsection{Decomposition into Lorentz Structures} \label{sec:BTT}
We begin by decomposing the matrix elements of the processes $M^{(*)} \bar{M}^{(*)} \to \gamma^*$ into invariant Lorentz structures, following the Bardeen--Tung--Tarrach (BTT) procedure~\cite{Bardeen:1968ebo,Tarrach:1975tu}. In this way, we obtain form factors as the scalar coefficient functions that are guaranteed to be free of kinematic constraints, singularities, and zeros, allowing us to set up a dispersive framework. As all of the processes are mediated by the strong interaction (and we disregard the QCD $\theta$-term), we only need to take into account structures that conserve parity.

For the pseudoscalar process $M \bar{M} \to \gamma^*$, we only have one independent Lorentz structure, resulting in the decomposition
\begin{equation}
	\bra{0} j^\mu(0) \ket{M(p_1)\,\bar{M}(p_2)} = r^\mu \, F_{M\bar{M}}\bigl(q^2\bigr)\,,
\end{equation}
where $r=p_1-p_2$. Due to the even parity of this reaction, the form factor $F_{M\bar{M}}\bigl(q^2\bigr)$ is of the electric type.

\begin{sloppypar}
For the vector--pseudoscalar transition $M^{*} \bar{M} \to \gamma^*$, we again only have one structure preserving the odd intrinsic parity,
\begin{align}
	\bra{0} &j^\mu(0) \ket{M^{*}(p_1,\lambda_1)\,\bar{M}(p_2)} \notag\\
    &= \epsilon^{\mu\nu\alpha\beta} \, q_\nu \, \epsilon_\alpha(\vec{p}_1,\lambda_1) \, p_{1\beta} \, F_{M^{*}\bar{M}}\bigl(q^2\bigr)\,,
\end{align}
giving us a magnetic-type transition form factor $F_{M^{*}\bar{M}}\bigl(q^2\bigr)$.
\end{sloppypar}

Finally, for the vector process $M^{*} \bar{M}^{*} \to \gamma^*$ we find three different structures. Following the conventions from the literature~\cite{Kim:1973ee,Arnold:1979cg,Arnold:1980zj,Brodsky:1992px}, they are
\begin{align}
	\bra{0} j^\mu(0) & \ket{M^{*}(p_1,\lambda_1)\,\bar{M}^{*}(p_2,\lambda_2)} \notag\\
    &= \epsilon_\alpha(\vec{p}_1,\lambda_1) \, \epsilon_\beta(\vec{p}_2,\lambda_2) \, \Gamma^{\mu\alpha\beta}(q,r)\,, \notag\\
\label{eq:VV_FF_decomposition}
	\Gamma^{\mu\alpha\beta}(q,r) &= r^\mu g^{\alpha\beta} \, F_1\bigl(q^2\bigr) + (q^\alpha g^{\mu\beta} - q^\beta g^{\mu\alpha}) \, F_2\bigl(q^2\bigr) \notag\\
    &+ \frac{r^\mu q^\alpha q^\beta}{2 \mV^2}\, F_3\bigl(q^2\bigr)\,,
\end{align}
where $\mV$ denotes the vector-meson mass. Note that there are two other common choices for the basis. The helicity basis can be obtained via the linear combinations\footnote{We use the set of polarization vectors found in \ref{app:polarization}.}
\begin{align} \label{eq:vectormeson_FF_helicity_basis}
	H_{11}\bigl(q^2\bigr) &= F_1\bigl(q^2\bigr)\,, \qquad H_{10}\bigl(q^2\bigr) = F_2\bigl(q^2\bigr)\,, \notag\\
	H_{00}\bigl(q^2\bigr) &= \frac{1}{2\mV^2} \biggl( \bigl(2\mV^2 - q^2\bigr) F_1\bigl(q^2\bigr) + q^2 F_2\bigl(q^2\bigr) \notag\\
    &\hspace{40pt}- \frac{1}{4\mV^2} q^2 \bigl(q^2-4\mV^2\bigr) F_3\bigl(q^2\bigr) \biggr)\,,
\end{align}
which, however, leads to the kinematic constraints
\begin{align}
	H_{00}(0) &= H_{11}(0)\,, \notag\\
    H_{00}\bigl(4\mV^2\bigr) &= - H_{11}\bigl(4\mV^2\bigr) + 2 H_{10}\bigl(4\mV^2\bigr)\,.
\end{align}
Also, one can relate these three form factors to the electric charge form factor $G_C$, the magnetic dipole form factor $G_M$, and the electric quadrupole form factor $G_Q$ via~\cite{Arnold:1979cg}
\begin{align}
	G_C &= F_1 + \frac{2}{3}\eta\, G_Q \approx F_1\,,\notag\\
    G_M &= F_2\,, \notag\\
	G_Q &= F_1 - F_2 + (1+\eta) F_3 \approx F_1 - F_2 + F_3\,,
\end{align}
with $\eta = -q^2/4\mV^2 \approx 0$ in the heavy-quark limit for small energies. Hence, $F_1$ can be associated with the electric form factor, $F_2$ with the magnetic dipole form factor, and all three contribute to the electric quadrupole form factor. Note that this again introduces a kinematic constraint,
\begin{equation}
    G_C\bigl(4\mV^2\bigr) = G_M\bigl(4\mV^2\bigr) + \frac{1}{3} G_Q\bigl(4\mV^2\bigr)\,.
\end{equation}

\subsection{Unitarity Relations for $\pi\pi$ Intermediate States}\label{sec:unitarity_relations}
Now that we have suitable BTT bases at hand, we can derive unitarity relations for the form factors for $\pi\pi$ intermediate states. For that purpose, we introduce the pion vector form factor $F_\pi^V(q^2)$,
\begin{equation}
    \bra{0} j^\mu(0) \ket{\pi^+(k_1)\,\pi^-(k_2)} = (k_1-k_2)^\mu F_\pi^V(q^2)\,,
\end{equation}
with $q=k_1+k_2$, as well as the $M^{(*)} \bar{M}^{(*)} \to \pi^+ \pi^-$ amplitudes,
\begin{align}
    \mathcal{M}&_{M^{(*)}\bar{M}^{(*)} \to \pi\pi}(s,t,u) \notag\\
    &\equiv\mathcal{M}\Bigl[M^{(*)}(p_1)\,\bar{M}^{(*)}(p_2) \to \pi^+(k_1)\, \pi^-(k_2)\Bigr]\,.
\end{align}
Here, we further define the Mandelstam variables
\begin{align}
    s&=(p_1+p_2)^2=(k_1+k_2)^2\,,\notag\\
    t&=(p_1-k_1)^2=(p_2-k_2)^2\,,\notag\\
    u&=(p_1-k_2)^2=(p_2-k_1)^2\,,
\end{align}
fulfilling $s+t+u=M_{P,V}^2+M_{\bar{P},\bar{V}}^2+2\mpi^2$, as well as the scattering angle
\begin{equation}
    z = \cos \theta = \frac{t-u}{\sigma_\pi(s) \lambda^{1/2}_{M^{(*)}\bar{M}^{(*)}}(s)}\,,
\end{equation}
using the short-hand notation $\sigma_\pi(s) = \sqrt{1-4\mpi^2/s}$ and $\lambda_{AB}(s) = \lambda\bigl(s,M_{A}^2,M_{B}^2\bigr)$.
We also denote the center-of-mass momenta as $\pcm = \sqrt{s}\,\sigma_\pi(s)/2$ and $\pz = \lambda^{1/2}_{M^{(*)}\bar{M}^{(*)}}(s)/2\sqrt{s}$ from now on.\footnote{Note that in order to have the correct analytic continuation, one needs to choose the square roots according to $\lambda^{1/2}_{AB}(s)=\sqrt{s-(M_{A}+M_{B})^2}\sqrt{s-(M_{A}-M_{B})^2}$.}

\begin{figure}
    \centering
    \includegraphics[width=0.3\linewidth]{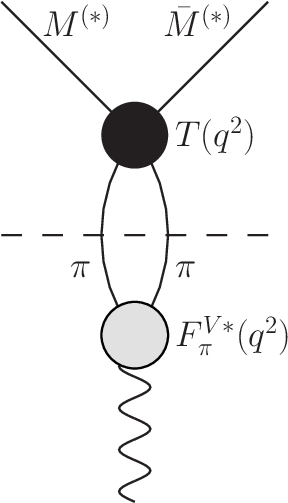}
    \caption{Diagrammatic representation of the form factor unitarity relation in Eq.~\eqref{eq:FF_unitarity}.}
    \label{fig:FF_unitarity}
\end{figure}

We find the following unitarity relation:
\begin{align}\label{eq:FF_unitarity}
	{\rm disc}_{\pi\pi}\, &\bra{0} j^\mu(0) \ket{M^{(*)}(p_1)\,\bar{M}^{(*)}(p_2)} \notag\\ 
	&= \frac{\ii}{4\pi^2} \left[F_\pi^V(q^2)\right]^* \int \dd^4\ell\, (q-2\ell)^\mu \notag\\
    &\times \mathcal{M}\Bigl[M^{(*)}(p_1)\,\bar{M}^{(*)}(p_2) \to \pi^+(q-\ell)\,\pi^-(\ell)\Bigr] \notag\\
    &\times\delta^{(+)}\bigl((q-\ell)^2-M_\pi^2\bigr) \, \delta^{(+)}\bigl(q^2-M_\pi^2\bigr) \,,
\end{align}
represented by the diagram in Fig.~\ref{fig:FF_unitarity}. We will drop the $\pi\pi$ subscript in the following, as we always consider the same discontinuity. In the next sections we will simplify this unitarity relation for the three different cases separately.

\subsubsection{Pseudoscalar Form Factor} \label{sec:unitarity_PFF}
In the pseudoscalar case, the $M\bar{M}\to\pi\pi$ scattering amplitude has even intrinsic parity and can be written as a simple scalar function,
\begin{equation}
    \mathcal{M}_{M\bar{M}\to\pi\pi}(s,t,u) = \mathcal{F}_{M\bar{M}}(s,z) \,,
\end{equation}
whose partial-wave expansion is given by~\cite{Jacob:1959at}
\begin{align}
    \mathcal{F}_{M\bar{M}}(s,z) &= \sum_{l=0}^{\infty} (2\ell+1) \, P_\ell(z) \, f_{M\bar{M}}^\ell(s)\,, \notag\\
	f_{M\bar{M}}^\ell(s) &= \frac{1}{2} \int_{-1}^{1} \dd{z}\, P_\ell(z) \, \mathcal{F}_{M\bar{M}}(s,z)\,,
\end{align}
where $P_\ell(z)$ are the Legendre polynomials. For later convenience, we rescale the $P$-wave amplitude as 
\begin{equation}
    T_{M\bar{M}}(s) = \frac{3}{2\pcm\pz} f_{M\bar{M}}^1(s) 
\end{equation}
to avoid kinematic zeros. The unitarity relation then reduces to the form
\begin{equation}
	\disc F_{M\bar{M}}(s) = 2\ii \,\theta\bigl(s-4\mpi^2\bigr) \frac{\pcm^3}{12\pi \sqrt{s}} T_{M\bar{M}}(s) \left[F_\pi^V(s)\right]^* .
\end{equation}

\subsubsection{Transition Form Factor} \label{sec:unitarity_TFF}
In the case of the vector--pseudoscalar transition form factor, the $M^*\bar{M}\to\pi\pi$ scattering amplitude has odd intrinsic parity and can therefore be expressed as 
\begin{align}
    \mathcal{M}_{M^*\bar{M}\to\pi\pi}(s,t,u) &= \epsilon_{\mu\nu\alpha\beta}  \, p_1^\alpha \, \epsilon^\beta(\vec{p}_1,\lambda_1) \notag\\
    &\times k_1^\mu \, k_2^\nu \, \mathcal{F}_{M^*\bar{M}}(s,z) 
\end{align}
via a scalar function $\mathcal{F}_{M^*\bar{M}}$. It has the partial-wave expansion~\cite{Jacob:1959at}
\begin{align} \label{eq:VPpipi_partial_waves}
	\mathcal{F}_{M^*\bar{M}}(s,z) &= \sum_{\ell=1}^{\infty} \, P^\prime_\ell(z) \, f_{M^*\bar{M}}^\ell(s) \,, \notag\\
	f_{M^*\bar{M}}^\ell(s) &= \frac{1}{2} \int_{-1}^{1} \dd{z} \big[P_{\ell-1}-P_{\ell+1}\big](z) \, \mathcal{F}_{M^*\bar{M}}(s,z) ,
\end{align}
where $P^\prime_\ell(z)$ denotes the derivatives of the Legendre polynomials.
Setting $T_{M^*\bar{M}}(s) = f_{M^*\bar{M}}^1(s)$, we can then bring the unitarity relation into the same form as in the pseudoscalar case,
\begin{align}
	\disc &F_{M^*\bar{M}}(s) \notag\\
    &= 2\ii \,\theta\bigl(s-4\mpi^2\bigr) \frac{\pcm^3}{12\pi \sqrt{s}} T_{M^*\bar{M}}(s) \left[F_\pi^V(s)\right]^* .
\end{align}

\subsubsection{Vector Form Factors} \label{sec:unitarity_VFF}
For the vector form factors, the $M^*\bar{M}^*\to\pi\pi$ amplitude has even intrinsic parity and can be decomposed into Lorentz structures 
\begin{align}
	\mathcal{M}_{M^*\bar{M}^*\to\pi\pi}&(s,t,u) \notag\\  &= \epsilon_\alpha(\vec{p}_1,\lambda_1) \, \epsilon_\beta(\vec{p}_2,\lambda_2) \, \tilde{\Gamma}^{\alpha\beta}(q,r,k) \,,
\end{align}
with $k = k_1 - k_2$, $k_\perp \equiv k - r \, (k \cdot r)/r^2 = k - r \, (z \pcm/\pz)$, and
\begin{align} 
	\tilde{\Gamma}^{\alpha\beta}&(q,r,k) \notag\\ &= A\, g^{\alpha\beta} + B\, q^\alpha q^\beta + C\, k_\perp^\alpha k_\perp^\beta + D\, k_\perp^\alpha q^\beta + E\, q^\alpha k_\perp^\beta\,.
    \label{eq:VVpipi_decomposition}
\end{align}
By charge-conjugation symmetry, we see that $E = -D$. We find the following four independent helicity amplitudes $\mathcal{F}_{\lambda_1 \lambda_2} = \epsilon_\alpha(\vec{p}_1,\lambda_1) \, \epsilon_\beta(\vec{p}_2,\lambda_2) \, \tilde{\Gamma}^{\alpha\beta}(q,r,k)$:
\begin{align} \label{eq:VVpipi_helicity_amplitudes}
	\mathcal{F}_{11}(s,\theta) &= A - 2 \pcm^2 \sin^2 \theta \, C \,, \notag\\
	\mathcal{F}_{10}(s,\theta) &= -\frac{\sqrt{2s} \pz \pcm \sin \theta}{\mV} D \,, \notag\\
	\mathcal{F}_{1,-1}(s,\theta) &= - 2 \pcm^2 \sin^2 \theta \, C\,, \notag\\
	\mathcal{F}_{00}(s,\theta) &= \frac{1}{2\mV^2} \left( \bigl(2\mV^2-s\bigr) \, A - 2s\pz^2 B \right)\,,
\end{align}
which have the partial-wave expansions~\cite{Jacob:1959at}
\begin{align} \label{eq:VVpipi_partial_waves}
	\mathcal{F}_{\lambda_1 \lambda_2}(s,\theta) &= \sum_{\ell=\abs{\lambda_1-\lambda_2}}^{\infty} (2\ell+1) \, d_{\abs{\lambda_1-\lambda_2},0}^\ell(\theta) \, f_{\lambda_1 \lambda_2}^\ell(s)\,, \notag\\
	f_{\lambda_1 \lambda_2}^\ell(s) &= \frac{1}{2} \int_{0}^{\pi} \dd{\theta} \sin \theta \, d_{\abs{\lambda_1-\lambda_2},0}^\ell(\theta) \, \mathcal{F}_{\lambda_1 \lambda_2}(s,\theta)\,,
\end{align}
where $d_{\lambda \lambda^\prime}^\ell(\theta)$ are the Wigner $d$-matrices that fulfill the orthogonality relation
\begin{equation}
	\int_{0}^{\pi} \dd{\theta} \sin \theta \, d_{\lambda \lambda^\prime}^\ell(\theta) \, d_{\lambda \lambda^\prime}^{\ell^\prime}(\theta) = \delta_{\ell\ell^\prime} \frac{2}{2\ell+1}\,.
\end{equation}
Note that we have the two special cases
\begin{align}
	d_{00}^\ell(\theta) &= P_\ell(\cos \theta)\,,\notag\\
    d_{10}^\ell(\theta) &= -\frac{1}{\sqrt{\ell(\ell+1)}} \sin \theta \, P_\ell^\prime(\cos \theta)\,.
\end{align}
For later purposes, we are only interested in the $P$-wave projections, which we rescale according to
\begin{align}
	g_{11}(s) &\equiv \frac{3f_{11}^1(s)}{2 \pz \pcm}\,,\quad\quad\quad
    g_{10}(s) \equiv -\frac{3 \mV f_{10}^1(s)}{\sqrt{s} \pz \pcm}\,,\notag\\ 
    g_{00}(s) &\equiv \frac{3 f_{00}^1(s)}{2 \pz \pcm}\,,
\end{align}
to eliminate kinematic zeros. Further, by setting
\begin{align}
	T_1(s) &\equiv g_{11}(s), \quad\qquad T_2(s) \equiv g_{10}(s), \notag\\
	T_3(s) &\equiv \frac{4\mV^2}{s\bigl(s-4\mV^2\bigr)} \bigl( (2\mV^2 - s) g_{11}(s) + s \,g_{10}(s) \notag\\
    &\hspace{70pt}- 2M_V^2 g_{00}(s) \bigr)\,,
\end{align}
we again find the same form for the form factor unitarity relations,
\begin{equation}
	\disc F_{i}(s) = 2\ii \,\theta\bigl(s-4\mpi^2\bigr) \frac{\pcm^3}{12\pi \sqrt{s}} T_{i}(s) \left[F_\pi^V(s)\right]^*\,.
\end{equation}
%

\section{\texorpdfstring{$D^{(*)} \bar D^{(*)} \to \pi\pi$}{D(*)D(*)→ππ} and \texorpdfstring{$B^{(*)} \bar B^{(*)}  \to \pi\pi$}{B(*)B(*)→ππ} \texorpdfstring{$P$}{P}-Wave Amplitudes}\label{sec:scattering}

\begin{figure}
    \centering
    \includegraphics[width=\linewidth]{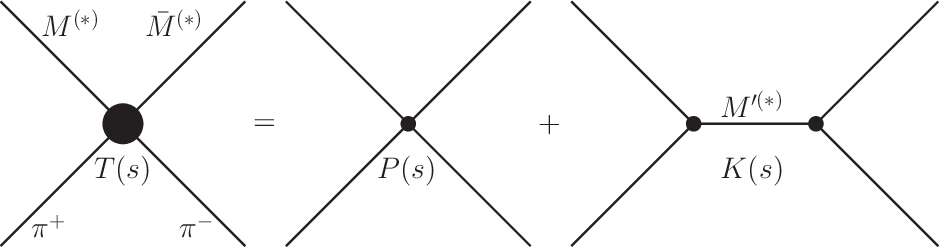}
    \caption{Diagrammatic representation of the amplitudes for $M^{(*)}\bar{M}^{(*)}\to\pi\pi$ as a sum of a contact term and an $M^{\prime(*)}$ left-hand cut.}
    \label{fig:LHC_DD}
\end{figure}

In the next step, we model the $M^{(*)}\bar{M}^{(*)}\to\pi\pi$ $P$-wave amplitudes $T_i(s)$ as the sum of (polynomial) contact terms $P_i(s)$ and the $P$-wave projected $t$- or $u$-channel Born exchange $K_i(s)$ of an additional pseudoscalar or vector meson $M^{\prime(*)}$,
\begin{equation}
    T_i(s) = P_i(s) + K_i(s)\,,
\end{equation}
leading to a left-hand cut; see Fig.~\ref{fig:LHC_DD}. In case of both the pseudoscalar and the transition form factors, only a vector-meson exchange can contribute, as a triple-pseudoscalar $M M^\prime \pi$ vertex is forbidden by parity within the strong interaction. In the vector case, both pseudoscalar and vector-meson exchanges contribute.

To model these terms, a combination of chiral perturbation theory (ChPT) and heavy-quark effective theory is used. The effective Lagrangian density at leading order (LO) for the interaction between heavy mesons and the Goldstone bosons is given by~\cite{Wise:1992hn}
\begin{align}
    \label{eq:HQET_Lagrangian}
    \mathcal{L}=& -\ii \, \tr[\bar{H}_a v^\mu\partial_\mu H_a] \notag\\ 
    &+\frac{\ii}{2} \,\tr[\bar{H}_a H_bv^\mu] \, (u^\dagger\partial_\mu u+u\partial_\mu u^\dagger)_{ba}  \notag\\ 
    & +\frac{\ii g}{2} \, \tr[\bar{H}_a H_b\gamma^\mu\gamma_5] \,(u^\dagger\partial_\mu u-u\partial_\mu u^\dagger)_{ba}+\dots\,.
\end{align}
Here, the Goldstone bosons are incorporated via the unitary matrix $u$, given by
\begin{align}
\label{eq:U-def}
	u&=\exp\left( \frac{\ii\pi}{2\Fpi}\right)\,, & 
	\pi&= 
	\begin{pmatrix}
	\pi^0 & \sqrt{2}\,\pi^+\\
	\sqrt{2}\,\pi^- & -\pi^0
	\end{pmatrix}\,,
\end{align}
and the pion decay constant $\Fpi=92.32(10)\MeV$~\cite{ParticleDataGroup:2024cfk}.\footnote{Note that $f_\pi = \sqrt{2} \Fpi$ to compare with the convention used in Ref.~\cite{ParticleDataGroup:2024cfk}.} The heavy mesons are encoded in the heavy superfields
\begin{align}
    \label{eq:H-fields}
    H_a &= \frac{1+\slashed{v}}{2}\left(P^*_{a\mu}\gamma^\mu+\ii P_a\gamma_5\right) \,,\notag\\
    \bar{H}_a &= \left(P^{*\dagger}_{a\mu}\gamma^\mu+\ii P^\dagger_a\gamma_5\right)\frac{1+\slashed{v}}{2}\,,
\end{align}
with four-velocity $v^\mu$ and
\begin{align}
	P^{(*)}=\left(P_1^{(*)},P_2^{(*)}\right)=\begin{cases}
	    \left(D^{(*)0},D^{(*)+}\right)\,,\\
	    \left(B^{(*)-},\bar{B}^{(*)0}\right)\,,
	\end{cases}
\end{align}
where each pseudoscalar (vector) meson field contains a normalization factor of $\mP^{1/2}$ ($\mV^{1/2}$) and where all vector mesons fulfill $P^*_{a,\mu} v^\mu = 0$.

The coupling constant $g$ is calculated from the partial decay widths of the decays $D^{*+}\rightarrow D^+\pi^0$ and $D^{*+}\rightarrow D^0\pi^+$~\cite{ParticleDataGroup:2024cfk} and translated to the $B$-meson case by heavy-quark symmetry. The result is given by
\begin{align}
\label{eq:g}
    |g|=0.547(6)\,.
\end{align} 

In the following, we will discuss several subtleties in the analytic structure of the form factors under investigation; \cf\ Sect.~\ref{sec:anomalous_thresholds} on anomalous thresholds.  For this reason, it is of utmost importance to maintain exact relativistic kinematics wherever possible, which is by no means guaranteed in a non-relativistic heavy-mass expansion (\cf\ also Refs.~\cite{Becher:1999he,Kubis:2000zd}). To obtain a  covariant interaction Lagrangian, we use the replacements 
\begin{align}\label{eq:HQET_relativistic_replacement}
    P^{(*)}_{a} &\mapsto M_{P,V}^{1/2} P^{(*)}_{a}\,,\notag\\
    P^{(*)\dagger}_{a} v^\mu P^{(*)}_{b} &\mapsto \frac{\ii}{2} \Bigl[ P^{(*)\dagger}_{a} \left(\partial^\mu P^{(*)}_{b}\right) - \left( \partial^\mu P^{(*)\dagger}_{a} \right) P^{(*)}_{b} \Bigr]\,,
\end{align}
resulting in the three-point interaction terms
\begin{align} \label{eq:lagrangian_L3}
    \mathcal{L}_{3}&=\frac{\ii g}{\Fpi} \sqrt{\mV\mP} \left[ P^{*\dagger\mu}_a P_b^{\phantom*} -P^{\dagger}_a P^{*\mu}_b \right] \partial_\mu \pi_{ba} \notag\\
    &+ \frac{g}{2\Fpi} \epsilon^{\alpha\beta\mu\nu} \Bigl[ P^{*\dagger}_{a,\alpha} \left( \partial_\mu P^{*}_{b,\beta} \right) - \left( \partial_\mu P^{*\dagger}_{a,\alpha} \right) P^{*}_{b,\beta} \Bigr] \partial_\nu \pi_{ba}\,,
\end{align}
as well as the Weinberg--Tomozawa~\cite{Weinberg:1966kf,Tomozawa:1966jm} four-point contact terms
\begin{align} \label{eq:lagrangian_L4}
    \mathcal{L}_{4}&=\frac{1}{8\Fpi^2} \big[\pi\, \big(\partial_\mu \pi\big)-\big(\partial_\mu \pi\big)\,\pi\big]_{ba} \notag\\
    &\times \Big(\Bigl[ P^{\dagger}_{a} \left( \partial^\mu P_{b}^{\phantom*} \right) - \left(\partial^\mu P^{\dagger}_{a} \right) P_{b}^{\phantom*} \Bigr] \notag\\
    &\hspace{10pt}- \Bigl[ P^{*\dagger}_{a,\alpha} \left( \partial^\mu P^{*\alpha}_{b} \right) -\left( \partial^\mu P^{*\dagger}_{a,\alpha} \right) P^{*\alpha}_{b} \Bigr]\Big)\,.
\end{align}
The resulting Feynman rules are summarized in~\ref{app:feynman_rules}.

We further need next-to-leading-order (NLO) four-point contact terms. We construct these by defining the chiral building blocks~\cite{Fettes:1998ud,Holmberg:2018dtv}
\begin{align}
    u_\mu & = \ii \big( u^\dagger (\partial_\mu - \ii r_\mu) u - u (\partial_\mu -\ii l_\mu)  u^\dagger \big)\,, \notag\\
    \chi_\pm &= u^\dagger \chi u^\dagger \pm u \chi^\dagger u \,, \notag\\
    f_\pm^{\mu\nu} &= u^\dagger f^{\mu\nu}_R u \pm u f^{\mu\nu}_L u^\dagger \,,
\end{align}
where
\begin{align}
    \chi &= 2B(s+\ii p) \,, \notag\\
    f^{\mu\nu}_R &= \partial^\mu r^\nu -\partial^\nu r^\mu - i [r^\mu,r^\nu] \,, & 
    r^\mu & = v^\mu + a^\mu \,, \notag\\
    f^{\mu\nu}_L &= \partial^\mu l^\nu -\partial^\nu l^\mu - i [l^\mu,l^\nu] \,, & 
    l^\mu & = v^\mu - a^\mu 
\end{align}
collect external scalar ($s$), pseudoscalar ($p$), vector ($v^\mu$), and axialvector ($a^\mu$) sources.
With these, we define the following complete and minimal NLO Lagrangian (see also Ref.~\cite{Jiang:2019hgs}): 
\begin{align} \label{eq:NLO_Lagrangian_nonrel}
    \mathcal{L}^\text{NLO} &= 
    c_1 \,\tr [\bar H_a H_a] \, (\chi_+)_{bb}  \notag\\
    &+ c_2 \,\tr[\bar H_a H_a] \, v^\mu v^\nu \, (u_\mu u_\nu)_{bb} \notag\\
    &+ c_3 \,\tr[\bar H_a H_a] \, (u_\mu u^\mu)_{bb} \notag\\
    &+ \ii  c_4 \tr[\bar H_a   H_b \sigma^{\mu\nu} ] \, [u_\mu,u_\nu]_{ba} \notag\\ 
    &+ c_5 \,\tr(\bar H_a  H_b) (\hat\chi_+)_{ba}  \notag\\
    &+ c_6 \,\tr[\bar H_a  H_b \sigma_{\mu\nu}] \, (\hat f_+^{\mu\nu})_{ba}
    \notag\\
    &+ c_7 \,\tr[\bar H_a H_a \sigma_{\mu\nu}] \, (f_+^{\mu\nu})_{bb}
  \,,
\end{align}
where $\sigma^{\mu\nu} = \frac{\ii}{2} [\gamma^\mu, \gamma^\nu]$, and we denote traceless objects (in flavor space) by $\hat A \equiv A - \frac{1}{2}(A)_{aa} \mathds{1}_2$.

It turns out that the only NLO term relevant for $\pi\pi$ $P$-wave interactions is the one $\sim c_4$. This can again be translated into a covariant Lagrangian via the replacement in Eq.~\eqref{eq:HQET_relativistic_replacement} to yield
\begin{align} \label{eq:NLO_Lagrangian_rel}
    \mathcal{L}_4^\text{NLO} &= -\frac{\ii c_4}{\Fpi^2} \epsilon^{\alpha\beta\mu\nu} \big(\partial_\mu \pi \, \partial_\nu \pi - \partial_\nu \pi \, \partial_\mu \pi\big)_{ba} \notag\\
    &\times\Bigl[ P^{*\dagger}_{a,\alpha} \big(\partial_\beta P^{\phantom*}_{b}\big) - \big(\partial_\beta P^{*\dagger}_{a,\alpha}\big) P^{\phantom*}_{b} \notag\\
    &\hspace{15pt}- P^{\dagger}_{a} \big(\partial_\beta P^{*}_{b,\alpha}\big) + \big(\partial_\beta P^{\dagger}_{a}\big) P^{*}_{b,\alpha} \Bigr] \notag\\
    &+ \frac{4 c_4 \mV}{\Fpi^2} \big(\partial^\mu \pi \, \partial^\nu \pi - \partial^\nu \pi \, \partial^\mu \pi\big)_{ba} P^{*\dagger}_{a,\mu} P^{*}_{b,\nu}\,.
\end{align}
Again, the resulting Feynman rules are summarized in~\ref{app:feynman_rules}. Ideally, the constant $c_4$ in the NLO interaction term should be found by a fit to data on pion--$M$-meson scattering. However, unfortunately, no such data are available: since $D^*$ mesons are unstable particles, it is difficult to obtain any scattering information. Refs.~\cite{Guo:2009ct,Lutz:2022enz} address subsequent efforts to analyze grid data, but these are not yet available for the correct partial wave. Instead, we will fix this constant later in Sect.~\ref{sec:contact_fixing}.

In the following, we will also want to test the effects of a next-to-next-to-leading-order (NNLO) contact term that contributes to the scattering amplitude $T_3(s)$ and, hence, to the form factor $F_3(s)$, which is of higher order in the low-energy expansion; \cf\ Eq.~\eqref{eq:VV_FF_decomposition}.  Such a contact interaction can be produced by the Lagrangian term
\begin{align}  
    \mathcal{L} &= d_{\rm NNLO}\, \tr(\bar{H}_a \dvec\partial_\mu H_b \sigma^{\mu\nu}) \,  [u_\nu, u_\alpha]_{ba} \, v^\alpha \,,
\end{align}
where $A \dvec \partial_\mu B \equiv A\partial_\mu B - (\partial_\mu A) B$.
Translated into covariant form and reduced to the four-point vertices, this corresponds to 
\begin{align}
   \mathcal{L} &=\frac{d_{\rm NNLO}}{F_\pi^2}\left(\partial^\nu\pi \,\partial^\alpha\pi-\partial^\alpha\pi\,\partial^\nu\pi\right)_{ba}
   \notag\\ & \times \big[\left(\partial^\mu P^{*\dagger}_{a,\nu}\right)\left(\partial_\alpha P^{*}_{b,\mu}\right) 
    -\left(\partial_\alpha\partial^\mu P_{a,\nu}^{*\dagger}\right)P_{b,\mu}^*  \notag\\
    &\qquad + P_{a,\mu}^{*\dagger}(\partial_\alpha\partial^\mu P^*_{b,\nu})-(\partial_\alpha P^{*\dagger}_{a,\mu})(\partial^\mu P^*_{b,\nu})
    \big] \,.
    \label{NNLO-contact}
\end{align}
  The size of the low-energy constant $d_{\rm NNLO}$ will only be estimated by dimensional analysis.

\subsection{Born and Contact Terms}\label{sec:born}
For later convenience, we introduce the short-hand notation\footnote{Keep in mind that depending on which of the three cases we consider, the center-of-mass momentum $\pz$ depends on a different combination of the heavy masses and, thus, has a slightly different analytic structure; however, we do not want to make the notation more cluttered.}
\begin{equation}
	\zV = \frac{\mV^2 + \pcm^2 + \pz^2}{2 \pz \pcm}, \qquad \zP = \frac{\mP^2 + \pcm^2 + \pz^2}{2 \pz \pcm},
\end{equation}
and
\begin{align}\label{eq:logarithm_combinations}
	L_1^{P,V}(s) &= \frac{1}{\pcm^2\pz^2}\,L_1(z_{P,V})\,, \notag\\
	L_2^{P,V}(s) &= \frac{1}{\pcm\pz}\,L_2(z_{P,V})\,, \notag\\
	L_3^{P,V}(s) &= \frac{1}{\pcm^2\pz^2}\,L_3(z_{P,V})\,,
\end{align}
with
\begin{align}\label{eq:logarithm_combinations_z}
	L_1(z) &= 2 + z \log\left(\frac{z-1}{z+1}\right)\,, \notag\\
	L_2(z) &= 2z + (z^2-1) \log\left(\frac{z-1}{z+1}\right)\,, \notag\\
	L_3(z) &= \frac23 + (z^2-1) \, L_1(z) = -\frac43 + z \, L_2(z)\,. 
\end{align}
These combinations $L_i^{P,V}(s)$ have the advantage that they are free of kinematic singularities and zeros, owing to the fact that they behave as $L_{1,3}(z) \asymp z^{-2}$ and $L_{2}(z) \asymp z^{-1}$ for $z \to \infty$ as well as $L_i(z) \to \mathrm{const.}$\ for $z \to 0$.

\subsubsection{Pseudoscalar Case} \label{sec:born_PFF}
The isovector $M\bar{M}\to\pi\pi$ amplitude is given by the sum of a pole term and a contact term
\begin{equation}
	\mathcal{M} = \mathcal{M}^{\text{pole}}_V + \mathcal{M}^{\text{cont}}_{\text{LO}}\,,
\end{equation}
provided by
\begin{align}
	\mathcal{M}^{\text{pole}}_V &\left[ M(p_1) \, \bar{M}(p_2) \to \pi^+(k_1) \, \pi^-(k_2) \right] \notag\\
	&= -\frac{g^2 \mP}{\Fpi^2 \mV} \Biggl[ -2z\pcm^3\pz \biggl(\frac{1}{t-\mV^2}+\frac{1}{u-\mV^2}\biggr) \notag\\
	&\hspace{20pt}+ \bigl(\mV^2 \mpi^2 + 2 \mV^2 \pcm^2 + \pcm^4 + z^2 \pcm^2\pz^2\bigr) \notag\\
    &\hspace{82pt}\times \biggl(\frac{1}{t-\mV^2}-\frac{1}{u-\mV^2}\biggr) \Biggr]\,, \notag\\
	\mathcal{M}^{\text{cont}}_{\text{LO}} &\left[ M(p_1) \, \bar{M}(p_2) \to \pi^+(k_1) \, \pi^-(k_2) \right] = \frac{\pcm\pz z}{\Fpi^2}\,.
\end{align}
After projecting out the $P$~wave, we obtain
\begin{align} \label{eq:PFF_born_term}
	K_{M\bar{M}}(s) = &-\frac{3g^2 \mP \mV}{16 \Fpi^2}  \Bigl( 2s - s_+^{PV}\Bigr) L_1^V(s)
\end{align}
with\footnote{Note that  $s=s_+^{PV}$ is one of the two logarithmic branch points of $L_1^V(s)$, which will be elaborated on in Sect.~\ref{sec:anomalous_thresholds}.}
\begin{align}
    s_+^{PV}=-\frac{1}{\mV^2}\lambda\bigl(\mP^2,\mV^2,\mpi^2\bigr)
\end{align}
and
\begin{equation} \label{eq:PFF_contact_term}
	P_{M\bar{M}}(s) = \frac{1}{2\Fpi^2} \biggl( 1 - \frac{g^2 \mP}{\mV} \biggr)\,.
\end{equation}
Note that the constant term $-(g^2\mP) / (2\mV\Fpi^2)$ stems from the projection of $\mathcal{M}^{\text{pole}}_V$ and was shifted from $K_{M\bar{M}}(s)$ to $P_{M\bar{M}}(s)$.\footnote{See Appendix B of Ref.~\cite{Kang:2013jaa} for further discussion.} This will be convenient later, as it ensures the high-energy behavior
\begin{equation}
	K_{M\bar{M}}(s) \asymp \frac{1}{s}\,.
\end{equation}

\subsubsection{Transition Case} \label{sec:born_TFF}
The isovector $M^*\bar{M}\to\pi\pi$ amplitude is given by the sum of a pole term and a contact term
\begin{equation}
	\mathcal{M} = \mathcal{M}^{\text{pole}}_V + \mathcal{M}^{\text{cont}}_{\text{NLO}}\,,
\end{equation}
provided by
\begin{align}
	\mathcal{M}^{\text{pole}}_V &\left[ M^*(p_1, \lambda_1) \, \bar{M}(p_2) \to \pi^+(k_1) \, \pi^-(k_2) \right] \notag\\
	&= -\frac{g^2 \sqrt{\mP \mV}}{\Fpi^2} \, \epsilon_{\alpha\beta\mu\nu} \, p_1^\alpha \, \epsilon^\beta(\vec{p}_1,\lambda_1) \, k_1^\mu \, k_2^\nu \notag\\
	&\hspace{60pt}\times \biggl(\frac{1}{t-\mV^2}+\frac{1}{u-\mV^2}\biggr)\,,\notag\\
	\mathcal{M}^{\text{cont}}_{\text{NLO}} &\left[ M^*(p_1, \lambda_1) \, \bar{M}(p_2) \to \pi^+(k_1) \, \pi^-(k_2) \right] \notag\\
    &= \frac{4 c_4}{\Fpi^2} \epsilon_{\alpha\beta\mu\nu} \, p_1^\alpha \, \epsilon^\beta(\vec{p}_1,\lambda_1) \, k_1^\mu \, k_2^\nu\,.
\end{align}
Projecting out the $P$~wave leads to
\begin{equation}
	K_{M^*\bar{M}}(s) = \frac{3g^2 \sqrt{\mP \mV}}{4 \Fpi^2} L_2^V(s) \,, \quad
\label{eq:TFF_contact_term}
	P_{M^*\bar{M}}(s) = \frac{4 c_4}{\Fpi^2}\,.
\end{equation}
Again, at high energies, the projected Born term falls off as
\begin{equation}
	K_{M^*\bar{M}}(s) \asymp \frac{1}{s}\,.
\end{equation}

\subsubsection{Vector Case} \label{sec:born_VFF}

The isovector $M^*\bar{M}^*\to\pi\pi$ amplitude is given by the sum of two pole terms and two contact terms
\begin{equation}
	\mathcal{M} = \mathcal{M}^{\text{pole}}_V + \mathcal{M}^{\text{pole}}_P + \mathcal{M}^{\text{cont}}_{\text{LO}} + \mathcal{M}^{\text{cont}}_{\text{NLO}}\,,
\end{equation}
provided by
\begin{align}
	\mathcal{M}^{\text{pole}}_V &\left[ M^*(p_1, \lambda_1) \, \bar{M}^*(p_2, \lambda_2) \to \pi^+(k_1) \, \pi^-(k_2) \right] \notag\\
    &= -\frac{g^2}{\Fpi^2} \epsilon_{\alpha\beta\mu\nu} \epsilon_1^\alpha p_1^\mu g^{\beta\lambda} \epsilon_{\sigma\lambda\gamma\delta} \epsilon_2^\sigma p_2^\gamma \notag\\
    &\times\biggl(\frac{k_1^\nu k_2^\delta}{t-\mV^2} - \frac{k_2^\nu k_1^\delta}{u-\mV^2}\biggr) \,, \notag\\
	\mathcal{M}^{\text{pole}}_P &\left[ M^*(p_1, \lambda_1) \, \bar{M}^*(p_2, \lambda_2) \to \pi^+(k_1) \, \pi^-(k_2) \right] \notag\\
    &= -\frac{\mP \mV g^2}{\Fpi^2} \epsilon_{1\mu} \epsilon_{2\nu} \biggl(\frac{k_1^\mu k_2^\nu}{t-\mP^2} - \frac{k_2^\mu k_1^\nu}{u-\mP^2}\biggr) \,, \notag\\
	\mathcal{M}^{\text{cont}}_{\text{LO}} &\left[ M^*(p_1, \lambda_1) \, \bar{M}^*(p_2, \lambda_2) \to \pi^+(k_1) \, \pi^-(k_2) \right] \notag\\
    &= \frac{\pcm\pz z}{\Fpi^2}\, \epsilon_{1\mu} \epsilon_2^\mu \,, \notag\\ 
	\mathcal{M}^{\text{cont}}_{\text{NLO}} &\left[ M^*(p_1, \lambda_1) \, \bar{M}^*(p_2, \lambda_2) \to \pi^+(k_1) \, \pi^-(k_2) \right] \notag\\
    &= \frac{4 c_4 \mV}{\Fpi^2} k_1^\mu k_2^\nu (\epsilon_{2\mu} \epsilon_{1\nu} - \epsilon_{1\mu} \epsilon_{2\nu})\,,
\end{align}
with the short-hand notation $\epsilon_1 \equiv \epsilon(\vec{p}_1, \lambda_1)$ and $\epsilon_2 \equiv \epsilon(\vec{p}_2, \lambda_2)$. Using the projections from Sect.~\ref{sec:unitarity_VFF}, we find the $P$-wave Born amplitudes
{\allowdisplaybreaks
\begin{align}
	K_1(s) &= -\frac{3g^2 \mV^2}{8 \Fpi^2} \bigg(  \frac{s}{2} L_1^V(s) - \pcm^2 L_3^V(s) \bigg) \notag\\
    &+ \frac{3g^2 \mP \mV}{8 \Fpi^2} \pcm^2 L_3^P(s)\,, \notag\\
	K_2(s) &= \frac{3g^2 \mV^2}{8 \Fpi^2} \bigg( L_2^V(s) - \pcm^2 L_3^V(s) \bigg) \notag\\
    &+\frac{3g^2 \mP \mV}{8 \Fpi^2} \bigg( L_2^P(s) - \pcm^2 L_3^P(s) \bigg)\,, \notag\\
	K_3(s) &= \frac{3g^2 \mV^4}{16 \Fpi^2 s \pz^2} \bigg(  s\big(s-2\mV^2\big) L_1^V(s) \notag\\
    &\hspace{10pt}+ 2s L_2^V(s) - 4\big(s+\mV^2\big) \pcm^2 L_3^V(s) \bigg) \notag\\
	&+ \frac{3g^2 \mP \mV^3}{16 \Fpi^2 s \pz^2} \bigg( 2s\mP^2 L_1^P(s) \notag\\
    &\hspace{10pt}+ 2s L_2^P(s) - 2\big(3s-2\mV^2\big) \pcm^2 L_3^P(s) \bigg)\,,
\end{align}
}%
and contact terms
\begin{equation} \label{eq:VFF_contact_terms}
	P_1(s) = \frac{1}{2\Fpi^2}\,, \quad P_2(s) = \frac{4 c_4 M_V}{\Fpi^2}\,, \quad P_3(s) = 0\,.
\end{equation}
The Born terms possess the high-energy behavior
\begin{equation}
	K_1(s) \asymp \frac{1}{s}\,, \quad K_2(s) \asymp \frac{1}{s}\,, \quad K_3(s) \asymp \frac{1}{s^2}\,.
\end{equation}
Note that, unlike all four other cases, the amplitude $K_3(s)$ falls quadratically with $s$.

\subsection{Enforcing Heavy-Quark Symmetry} \label{sec:heavy_quark_symmetry}
In the heavy-quark limit, we expect the heavy-quark symmetry to be fulfilled by the amplitudes, meaning that both $T_{M\bar{M}}(s)$ and $T_1(s)$ should equal each other as well as $T_{M^*\bar{M}}(s)$ and $T_2(s)$. It is rather straightforward to see that the latter is fulfilled: the functions $L_3^{P,V}(s)$ are suppressed by one power of the heavy mass compared to $L_2^{P,V}(s)$, meaning they do not matter in this limit, thus leaving us with the same expressions overall. 

For the former two, this is also the case, but in a much more subtle way. A key difference is the lack of the constant term in $K_1(s)$ that we found in $K_{M\bar{M}}(s)$ and shifted to $P_{M\bar{M}}(s)$ [\cf\ Eq.~\eqref{eq:PFF_contact_term}]. While this does not matter at tree level as it is reproduced asymptotically in the heavy-quark limit, it would break the heavy-quark symmetry during the unitarization procedure in the following section. However, we make the following observation:
\begin{align}
	\pz^2\pcm^2 \, L^{V,P}_3(s) = \frac23 + \frac{M_{V,P}^2}{4} \Bigl(s-s_+^{V(V,P)}\Bigr) \, L^{V,P}_1(s)
\end{align}
with\footnote{Again, $s_+^{VV}$ and $s_+^{VP}$ are the branch points of the respective logarithms $L_i^{P,V}(s)$.}
\begin{align}
    s_+^{VV}&=-\frac{1}{\mV^2}\lambda\bigl(\mV^2,\mV^2,\mpi^2\bigr)\,, \notag\\
    s_+^{VP}&=-\frac{1}{\mP^2}\lambda\bigl(\mV^2,\mP^2,\mpi^2\bigr)\,.
\end{align}
Using this correspondence, we find
\begin{align} \label{eq:VFF_born_contact_symmetrized}
	K_1(s) &= -\frac{3g^2\mV^2}{32 \Fpi^2} \Big( 3s - s_+^{VV}  \Big) \, L^V_1(s) \notag\\
    &\phantom{=} -\frac{3g^2 \mP^3}{32 \Fpi^2 \mV} \Big( s - s_+^{VP}  \Big) \, L^P_1(s) \,, \notag\\
	P_1(s) &= \frac{1}{2\Fpi^2} \Biggl( 1 - \frac{g^2 (\mV+\mP)}{2\mV} \Biggr) \,,
\end{align}
upon neglecting the two terms
\begin{align}
    \sim \frac{3g^2}{32 \Fpi^2} s \pcm^2 L_3^V(s) + \frac{3g^2 M_P}{32 \Fpi^2 M_V} s \pcm^2 L_3^P(s)\,,
\end{align}
which in the low-energy region of interest are both suppressed chirally by a factor of $\pcm^2$ and via the heavy-mass scale by a factor $s/4M^2$ compared to all the other terms. Ignoring these terms is thus well justified, since we also neglected other corrections that are similarly suppressed (such as the difference between the $DD^*\pi$ and the $D^*D^*\pi$ couplings). Furthermore, the terms are composed of one of the singularity-free logarithms from Eq.~\eqref{eq:logarithm_combinations} and, thus, neglecting them does not alter the analytic structure. In doing so, we ensure heavy-quark symmetry is fulfilled, even after unitarization. This becomes evident by comparing Eq.~\eqref{eq:VFF_born_contact_symmetrized} with Eqs.~\eqref{eq:PFF_born_term} and~\eqref{eq:PFF_contact_term}.

\subsection{Unitarization} \label{sec:unitarization}
\begin{figure}
    \centering
    \includegraphics[width=\linewidth]{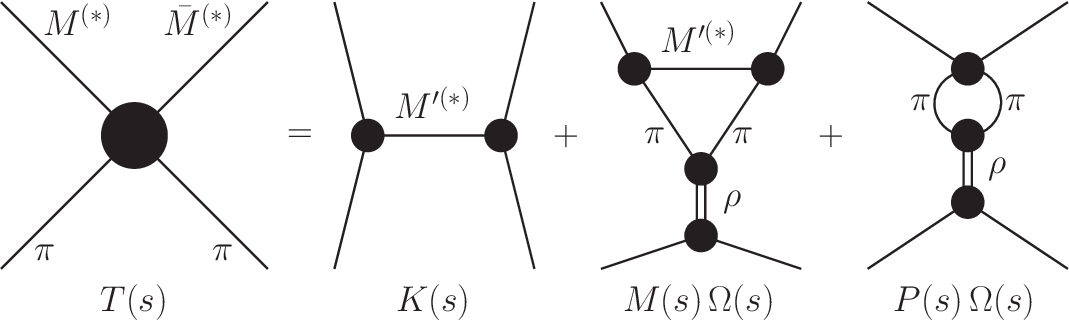}
    \caption{Diagrammatic representation of the unitarized $M^{(*)}\bar{M}^{(*)}\to\pi\pi$ $P$-wave amplitudes $T(s)$ as the sum of the Born term $K(s)$, the MO term $M(s)\,\Omega(s)$, and the subtraction term $P(s)\,\Omega(s)$. Notice the triangle topology in the MO term.}
    \label{fig:MO_sum_DD}
\end{figure}

So far, during the derivation of the $M^{(*)}\bar{M}^{(*)}\to\pi\pi$ $P$-wave amplitudes, we have just considered leading-order left-hand cut contributions plus leading polynomials, but completely ignored the possibility of final-state rescattering. Unitarity dictates, via Watson's theorem~\cite{Watson:1954uc}, that the phase of a $\pi\pi$ production partial wave (in the pseudo-elastic region) equals the corresponding $\pi\pi$ scattering phase shift, here $\delta(s) \equiv \delta_1^1(s)$. However, so far, the amplitudes do not match this---in fact, they do not have a right-hand cut at all. To resolve this issue, we unitarize the amplitudes via the Muskhelishvili--Omnès (MO) representation~\cite{Muskhelisvili:1953,Omnes:1958hv},
\begin{align} \label{eq:mo_solution}
	T_i(s) &= K_i(s) + \Omega(s) \left[ P_i(s) + M_i(s) \right]\,,\notag\\
    M_i(s)&=\frac{s}{\pi} \int_{4M_\pi^2}^{\infty} \frac{\dd{s^\prime}}{s^\prime} \frac{K_i(s^\prime) \sin \delta(s^\prime)}{(s^\prime-s) \, \abs{\Omega(s^\prime)}}\,.
\end{align}
It contains three contributions, represented by Fig.~\ref{fig:MO_sum_DD}, and is built from the Omnès function~\cite{Omnes:1958hv}
\begin{equation}
    \Omega(s) = \exp\left[\frac{s}{\pi}\int_{4M_\pi^2}^{\infty}\frac{\dd{s^\prime}}{s^\prime}\frac{\delta(s^\prime)}{s^\prime-s}\right]\,,
\end{equation}
which is the normalized solution to the homogeneous MO problem,
\begin{equation}
    \disc \Omega(s) = 2\ii\,\theta\big(s-4\mpi^2\big)\,\Omega(s)\,e^{-\ii\delta(s)}\sin\delta(s) \,,
\end{equation}
with $\Omega(0)=1$ and the once-subtracted MO integral $M_i(s)$ such that $T_i(s)$ solves the inhomogeneous MO problem 
\begin{align}
    \disc \bigl[T_i-K_i\bigr](s) = 2\ii\,\theta\big(s-4\mpi^2\big)\, T_i(s) \, e^{-\ii\delta(s)}\sin\delta(s)\,.
\end{align}
Note that one subtraction is sufficient, as the Omnès function behaves as $\Omega(s) \asymp s^{-1}$ and also all Born amplitudes fall at least as $K_i(s) \asymp s^{-1}$. The subtraction constants are provided by the (constant) contact terms $P_i(s)$ via matching at $s=0$. In the case of $T_3(s)$ even an unsubtracted dispersion relation would suffice, given that $K_3(s)\asymp s^{-2}$. Equivalently, one can still use a once-subtracted dispersion relation as for the others, and fix the contact term via a sum rule,
\begin{equation}
    P_3(s) = \frac{1}{\pi} \int_{4M_\pi^2}^{\infty} \dd{s^\prime}\, \frac{K_3(s^\prime) \sin \delta(s^\prime)}{s^\prime \, \abs{\Omega(s^\prime)}}\,.
\end{equation}

\subsection[Anomalous Thresholds]{Anomalous Thresholds\footnote{This section only contains the most relevant basics about anomalous thresholds. A more detailed discussion can be found in Sect.~2 of Ref.~\cite{Mutke:2024tww}; \cf\ also Refs.~\cite{Lucha:2006vc,Hoferichter:2013ama,Colangelo:2015ama,Hoferichter:2019nlq,Junker:2019vvy}.}} \label{sec:anomalous_thresholds}
With unitarization via the MO procedure, we ensure that the $M^{(*)}\bar{M}^{(*)}\to\pi\pi$ $P$-wave amplitudes have the correct two-particle cut structure mandated by unitarity, starting at the two-particle threshold $s_\text{thr}=4\mpi^2$. However, in a triangle topology, as it enters the MO representation, \cf\ Fig.~\ref{fig:MO_sum_DD}, there is the additional possibility of kinematic singularities arising when not only two, but all three internal particles of the triangle loop can go on-shell simultaneously. These so-called triangle singularities are of logarithmic nature and, thus, lead to additional branch cuts. While these only lie on the second Riemann sheet in many cases and are therefore not relevant for dispersive representations, they can actually move to the first Riemann sheet under certain circumstances and lead to additional \emph{anomalous thresholds} one needs to take care of.

\begin{figure}
    \centering
    \includegraphics[width=0.35\linewidth]{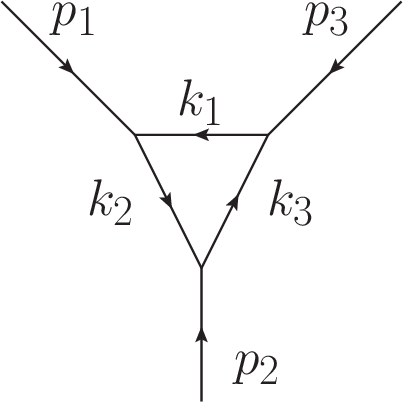}
    \caption{General triangle loop diagram.}
    \label{fig:loop_triangle}
\end{figure}

To analyze where these triangle singularities lie in the complex $s$-plane and under which circumstances they move to the first Riemann sheet, one can resort to Landau's equations~\cite{Landau:1959fi},
\begin{align}
    \sum_j \pm \alpha_j k_j = 0 \quad&\text{for each loop,}\notag\\
    \alpha_i=0~~\text{or}~~k_i^2=m_i^2 \quad&\text{for each progagator,}
\end{align}
determining the kinematic singularities of any given loop diagram with internal particles of mass $m_j$ and momentum $k_j$. A solution with all Feynman parameters $\alpha_i \neq 0$, and therefore all internal particles being on-shell, is called a leading Landau singularity, whereas solutions with at least one $\alpha_i=0$ are subleading singularities. The triangle singularities are the leading Landau singularities of triangle graphs as in Fig.~\ref{fig:loop_triangle}, \ie, the solutions of
\begin{equation}
	\begin{Bmatrix}
		0 = \alpha_1 k_1 + \alpha_2 k_2 + \alpha_3 k_3 \\[1mm]
		0 = k_1^2 - m_1^2 = k_2^2 - m_2^2 = k_3^2 - m_3^2
	\end{Bmatrix}\,.
\end{equation}
Setting $s=p_2^2$, these are found to lie at
\begin{align}\label{eq:triangle_singularities}
	s_\pm&=p_1^2 \frac{m_1^2 + m_3^2}{2m_1^2} + p_3^2 \frac{m_1^2+m_2^2}{2m_1^2} - \frac{p_1^2 p_3^2}{2m_1^2} \notag\\
    &- \frac{(m_1^2-m_2^2)(m_1^2-m_3^2)}{2m_1^2} \notag\\
	&\mp \frac{1}{2m_1^2} \lambda^{1/2}(p_1^2,m_1^2,m_2^2) \, \lambda^{1/2}(p_3^2,m_1^2,m_3^2)\,,
\end{align}
which coincides with singularities of the logarithms appearing in the partial-wave projected Born terms from Sect.~\ref{sec:born}. Landau's analysis further yields a criterion for one of the singularities to be on the first Riemann sheet. This is the case precisely when all Feynman parameters $\alpha_i$ corresponding to the solution of Landau's equations are strictly positive. In the case of the triangle singularities, this can never happen for $s_-$, but it can for $s_+$ if and only if
\begin{equation}\label{eq:triangle_singularity_condition}
	m_3 p_1^2 + m_2 p_3^2 > (m_2 + m_3)(m_1^2 + m_2 m_3).
\end{equation}
Depending on the signs of the two Källén functions $\lambda(p_1^2,m_1^2,m_2^2)$ and $\lambda(p_3^2,m_1^2,m_3^2)$, it can lie either on the real axis---both below or on the unitarity cut---or in the lower half-plane. 

\begin{figure}
    \centering
    \includegraphics[width=\linewidth]{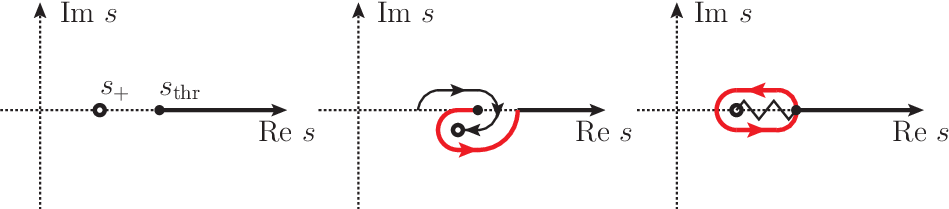}
    \caption{Left: Positions of $s_\text{thr}$ and $s_+$ in the complex $s$-plane before the analytic continuation in $\mV^2$. The integration path is indicated by the thick solid line. Middle: The path traced by $s_+$ during the analytic continuation is indicated by the thin solid line with arrows. When $s_+$ moves through the unitarity cut, the integration path needs to be deformed, picking up the additional red integration path. Right: Deformed integration path after the analytic continuation enclosing the anomalous branch cut, depicted by the zig-zag line.}
    \label{fig:anom_int_path}
\end{figure}

In our cases of interest, namely the $M^{(*)} \bar{M}^{(*)}$ form factors with $M=D,B$, it turns out that the triangle singularity $s_+$ only appears on the first sheet in the vector case with a pseudoscalar left-hand cut, such that $p_1^2=p_3^2=\mV^2$, $m_1^2=\mP^2$, and $m_2^2=m_3^2=\mpi^2$. This simplifies the kinematics a lot. The position of the triangle singularity in this case is
\begin{equation}\label{eq:triangle_position}
	s_+= s_+^{VP} =-\frac{1}{\mP^2} \lambda(\mV^2,\mP^2,\mpi^2)
\end{equation}
and it moves onto the first Riemann sheet as soon as $\mV^2 > \mP^2 + \mpi^2$. To see how this crossing happens, one can start with a lower value for the mass $\mV^2$, such that $\mV^2 < \mP^2 + \mpi^2$ and, thus, the triangle singularity is still on the second sheet, and then analytically continue in the mass back towards its physical value. As it turns out, in order to preserve the correct analytic structure during this process that agrees with causality, one needs to introduce an infinitesimal imaginary part to the vector-meson mass, \ie, $\mV^2 \mapsto \mV^2+\ii\delta$~\cite{Gribov:1962fu,Bronzan:1963mby}.\footnote{As soon as $\mV^2>(\mP+\mpi)^2$, a branch cut opens in the (complex) variable $\mV^2$, corresponding to $M^*$ becoming unstable, \ie,\ $M^* \to M \pi$ being allowed. Choosing the other branch $\mV^2 \mapsto \mV^2-\ii\delta$ would lead to a conflict with causality.} Close to the crossing point $\mV^2 = \mP^2 + \mpi^2$ the triangle singularity $s_+(\mV^2+\ii\delta)$ then traces the path displayed in Fig.~\ref{fig:anom_int_path}, which crosses the unitarity cut at
\begin{equation}
    s_\text{cross} = 4\mpi^2 + \frac{\delta^2}{\mP^2} + \mathcal{O}(\delta^4) > 4\mpi^2 = s_\text{thr}
\end{equation}
onto the first sheet and then moves along the real axis below the two-particle threshold $s_\text{thr}$. 

For $M=D$ we find $s_+\simeq-0.02\GeV^2$, which is at a small scale compared to $s_\text{thr}\simeq 0.078\GeV^2$ due to $M_{D^*}\gtrsim M_D+\mpi$, as evident from Eq.~\eqref{eq:triangle_position}. For $M=B$, on the other hand, we find $s_+\simeq 0.070\GeV^2$, which is comparatively close to threshold.

The consequence for the dispersive MO representation is that when the triangle singularity crosses the unitarity cut, and therefore the integration path, we need to deform the integration path around it as illustrated by Fig.~\ref{fig:anom_int_path}, picking up an additional anomalous integral,
\begin{align}
    M_i(s)&=\frac{s}{\pi} \int_{4M_\pi^2}^{\infty} \frac{\dd{s^\prime}}{s^\prime} \frac{K_i(s^\prime) \sin \delta(s^\prime)}{(s^\prime-s) \, \abs{\Omega(s^\prime)}} \notag\\
    &+ \frac{s}{\pi} \int_{0}^{1} \frac{\dd{x}}{s_x} \frac{\partial s_x}{\partial x} \frac{\disca K_i(s_x) \, \sigma_\pi(s_x) \, t_1^1(s_x)}{(s_x-s) \, \Omega(s_x)} \,,
\end{align}
which runs from the anomalous threshold $s_{x=0}=s_+$ to the usual two-particle threshold $s_{x=1}=s_\text{thr}$. Here, we used the isovector $P$-wave $\pi\pi$ scattering amplitude 
\begin{equation}
    t_1^1(s) = \frac{1}{\sigma_\pi(s)} e^{\ii\delta(s)} \sin \delta(s)
\end{equation}
to make the replacement~\cite{Hoferichter:2013ama,Niehus:2021iin}
\begin{equation}
    \frac{\sin \delta(s)}{\abs{\Omega(s)}} = \frac{\sigma_\pi(s) \, t_1^1(s)}{\Omega(s)}
\end{equation}
in the integrand, ensuring a proper analytic continuation away from the unitarity cut. The anomalous discontinuity $\disca K_i(s)$ is the discontinuity of $K_i(s)$ along its logarithmic cut and can be computed using the fact that $\disc(\log) = 2\pi\ii$. In particular, for the combinations of logarithms in Eq.~\eqref{eq:logarithm_combinations_z}, one has
\begin{align}
	\disca L_1(z) &= 2\pi\ii \, z\,, \notag\\
	\disca L_2(z) &= 2\pi\ii \left(z^2-1\right)\,, \notag\\
	\disca L_3(z) &= 2\pi\ii \, z \left(z^2-1\right)\,, 
\end{align}
allowing us to easily identify
\begin{align} \label{eq:born_disc_anom}
    \disca K_1(s) &= 2\pi\ii \frac{3g^2 \mP^3}{32 \Fpi^2 \mV \pcm^2 \pz^2} \, \Big(s_+^{VP}-s\Big)\, \zP \,, \notag\\
	\disca K_2(s) &=2\pi\ii \frac{3g^2 \mP \mV}{8 \Fpi^2 \pz^2 \pcm} \big(\pz -\pcm\zP\big)\big(\zP^2-1\big)\,, \notag\\
	\disca K_3(s) &= 2\pi\ii \frac{3g^2 \mP \mV^3}{8 \Fpi^2 s \pz^4 \pcm^2} \notag\\
    &\hspace{10pt}\times \Big( s\mP^2\,\zP + s\pcm\pz \big(\zP^2-1\big) \notag\\
    &\hspace{20pt}- \big(3s-2\mV^2\big) \pcm^2 \,\zP \big(\zP^2-1\big) \Big)\,.
\end{align}

In addition to adding the anomalous part to the MO representation as soon as $s_+$ enters the first sheet, one also needs to be careful about properly analytically continuing the logarithms inside the original dispersion integral. As it turns out, the logarithms can go to their second sheets, requiring the following replacement:
\begin{align} \label{eq:log_replacement_real}
    z &\log\left(\frac{z-1}{z+1}\right) \notag\\
    &\mapsto
    \begin{cases}
        z \log\left(\frac{z-1}{z+1}\right)\,,& \Im z = 0\,,\\[2mm]
        -2\abs{z}\left[ \arctan\left(\frac{1}{\abs{z}}\right) \right]\,,& \Im z < 0\,,\\[2mm]
        -2\abs{z}\left[ \arctan\left(\frac{1}{\abs{z}}\right) + \pi \right]\,,& \Im z > 0\,.
    \end{cases}
\end{align}
More details can be found in \ref{app:analytic_structure_partial_waves}.

An additional feature to note is that in the case $p_1^2=p_3^2=m_1^2=\mV^2$ and $m_2^2=m_3^2=\mpi^2$, the anomalous threshold lies at
\begin{equation}
    s_+ = s_+^{VV} = 4\mpi^2 - \frac{\mpi^4}{\mV^2}
\end{equation}
on the second Riemann sheet, very close to the $\pi\pi$ threshold. While this obviously does not add an analytic structure on the first sheet and, thus, does not require an anomalous contribution, it can, however, lead to an enhancement above the $\pi\pi$ threshold on the first sheet due to the close proximity of the triangle singularity. This phenomenon has also been observed in nucleon form factors~\cite{Frazer:1960zza,Hohler:1974eq,Bernard:1996cc}.

\section{Form Factor Dispersion Relations}\label{sec:formfactors}
\begin{figure}
    \centering
    \includegraphics[width=\linewidth]{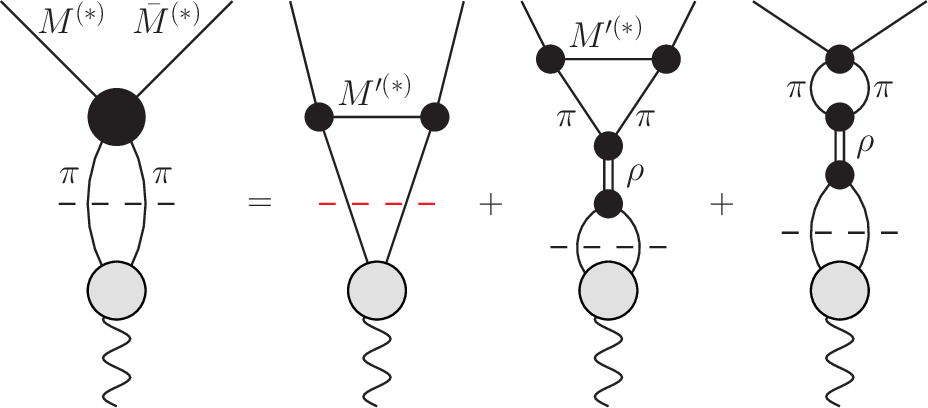}
    \caption{Diagrammatic representation of the form factor unitarity relation in Eq.~\eqref{eq:FF_unitarity_common_form} for $\pi\pi$ intermediate states where the unitarized $M^{(*)}\bar{M}^{(*)}\to\pi\pi$ amplitude is split up into its different contributions as in Fig.~\ref{fig:MO_sum_DD}. The red cut highlights that only the Born term leads to a (cut) triangle topology and, thus, an anomalous contribution.}
    \label{fig:FF_unitarity_sum}
\end{figure}

With the unitarized $M^{(*)}\bar{M}^{(*)}\to\pi\pi$ $P$-wave amplitudes $T_i(s)$ at hand, we now turn our attention to the $M^{(*)}\bar{M}^{(*)}\to\gamma^*$ form factors $F_i(s)$ themselves. The unitarity relations we derived in Sect.~\ref{sec:unitarity_relations} all share the common form
\begin{equation} \label{eq:FF_unitarity_common_form}
	\disc F_{i}(s) = 2\ii \,\theta\bigl(s-4\mpi^2\bigr) \frac{\pcm^3}{12\pi \sqrt{s}} T_{i}(s) \left[F_\pi^V(s)\right]^*\,.
\end{equation}
As both $F_\pi^V(s) \asymp s^{-1}$ and $T_{i}(s) \asymp s^{-1}$, the discontinuity falls as $\disc F_{i}(s) \asymp s^{-1}$ as well for large $s$ and we can, therefore, set up a once-subtracted form factor dispersion relation,
\begin{align} \label{eq:FF_dispersion_relation}
    F_i(s) &= F_i(0) + \frac{s}{12\pi^2} \int_{4\mpi^2}^{\infty} \frac{\dd s^\prime}{s^\prime} \frac{\pcm^3(s^\prime) \, T_i(s^\prime)\left[F_\pi^V(s^\prime)\right]^*}{s^{\prime 1/2} (s^\prime-s)}\notag\\
    \Bigg(&+ \frac{s}{12\pi^2} \int_{0}^{1} \frac{\dd x}{s_x} \frac{\partial s_x}{\partial x} \frac{\pcm^3(s_x) \, \disca K_i(s_x) F_\pi^V(s_x)}{s_x^{1/2} (s_x-s)}\Bigg),
\end{align}
whose subtraction constant is determined by the sum rule
\begin{align} \label{eq:FF_sum_rule}
    F_i(0) &= \frac{1}{12\pi^2} \int_{4\mpi^2}^{\infty} \dd s^\prime \frac{\pcm^3(s^\prime) \, T_i(s^\prime) \left[F_\pi^V(s^\prime)\right]^*}{s^{\prime 3/2}}\notag\\
    \Bigg(&+ \frac{1}{12\pi^2} \int_{0}^{1} \dd x \frac{\partial s_x}{\partial x} \frac{\pcm^3(s_x) \, \disca K_i(s_x) F_\pi^V(s_x)}{s_x^{3/2}}\Bigg).
\end{align}
In the cases where an anomalous threshold appears on the first sheet, one again needs to add an anomalous part to these integrals, as indicated inside the brackets. There are two observations to be made. First, the anomalous discontinuity is again only made up by the anomalous discontinuity of the Born term, $\disca K_i(s)$, since only the Born contribution of $T_i(s)$ leads to another triangle topology; see Fig.~\ref{fig:FF_unitarity_sum}. 
Second, in the anomalous integrand, compared to the integrand of the original integral, the pion vector form factor should not be complex conjugated anymore in order to have the proper analytic continuation into the complex plane~\cite{Mutke:2024tww}.

\subsection{Phase Input} \label{sec:phase_input}

\begin{figure}
    \centering
    \includegraphics[width=\linewidth]{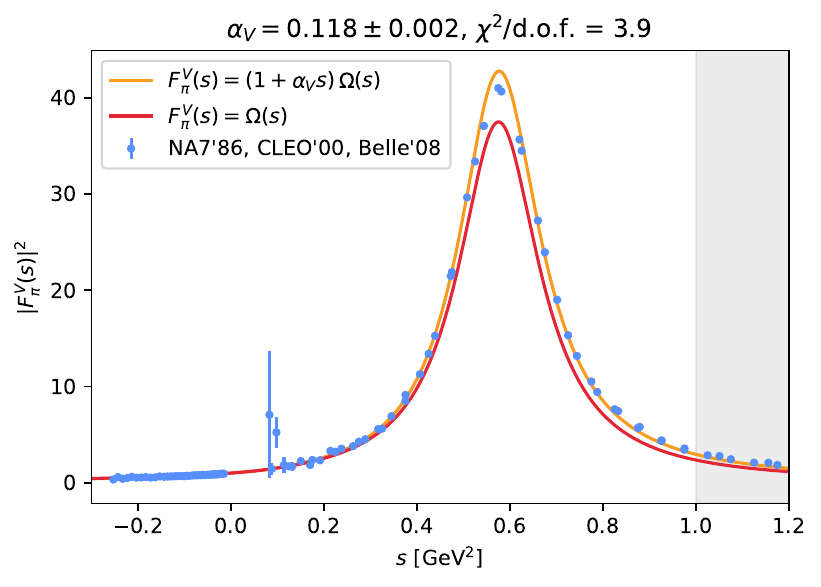}
    \caption{Determination of the slope parameter $\alpha_V$ of Eq.~\eqref{eq:pion_vff_polynomial} via a fit to $|F_\pi^V(s)|^2$ data from the CLEO, Belle, and NA7 experiments. For comparison the plain Omnès function $\Omega(s)$ is also shown.}
    \label{fig:pionVFF_slopeparameter_fit}
\end{figure}

One of the key ingredients of our formalism is the $\pi\pi$ isovector $P$-wave phase shift $\delta(s)$, which is needed as input to compute the Omnès function $\Omega(s)$ as well as both the MO and form factor dispersion relations. In the case of the presence of an anomalous threshold, we need in addition an analytic continuation of the $\pi\pi$ isovector $P$-wave amplitude $t_1^1(s)$ into the complex plane. In order not to violate unitarity, the phase of this amplitude along the real axis should match the phase shift $\delta(s)$ we use. We therefore use NLO ChPT results~\cite{Gasser:1983yg,Dax:2018rvs,Niehus:2020gmf},
\begin{equation}
    t_\text{ChPT}(s) = t_2(s) + t_4(s) + \mathcal{O}(p^6)\,,
\end{equation}
with
\begin{align}
t_2(s)&=\frac{s\sigma^2(s)}{96\pi F_0^2}\,, \notag\\
t_4(s)&=\frac{t_2(s)}{48\pi^2F_0^2}\bigg[s\bigg( \bar{l}+\frac{1}{3}\bigg)-\frac{15}{2}M_\pi^2-\frac{M_\pi^4}{2s}\bigg( 41-2L_\sigma \notag\\
&\times\big(73-25\sigma^2(s)\big)+3L_\sigma^2\big(5-32\sigma^2(s)+3\sigma^4(s)\big)\bigg) \bigg] \notag\\
&+\text{i}\sigma(s)\,[t_2(s)]^2, \notag\\
L_\sigma&=\frac{1}{\sigma^2}\left( \frac{1}{2\sigma(s)}\log\left[\frac{1+\sigma(s)}{1-\sigma(s)}\right]-1 \right)\,, 
\end{align}
that are unitarized via the inverse-amplitude method (IAM)~\cite{Dobado:1989qm,Dobado:1996ps},
\begin{equation}
    t_\text{IAM}(s) = \frac{[t_2(s)]^2}{t_2(s)-t_4(s)}\,,
\end{equation}
to model $t_1^1(s)$ and use its phase $\delta(s)=\arg\bigl[t_\text{IAM}(s)\bigr]$ as input. To reproduce the desired high-energy behavior of the Omnès function $\Omega(s) \asymp s^{-1}$, we further include some additional NNLO terms in $t_4(s)$~\cite{Holz:2015tcg,Holz:2024diw}
\begin{equation}
    t_4(s) \mapsto t_4(s) + \frac{t_2(s)}{48\pi^2 F_0^2} \Bigl( \hat{l}_s s^2 + \hat{l}_\pi \mpi^4 \Bigr)\,,
\end{equation}
such that $\delta(s)\to\pi$ for large values of $s$. The relevant constants are $\Fpi/F_0 = 1.062(7)$~\cite{FlavourLatticeAveragingGroupFLAG:2021npn,MILC:2010hzw,Borsanyi:2012zv,BMW:2013fzj,Boyle:2015exm,Beane:2011zm}, $\bar{l}=4.47(3)$, $\hat{l}_s=1.74(3)$, and $\hat{l}_\pi=560(20)$~\cite{Holz:2024diw}.
For the pion vector form factor we then use the parametrization
\begin{equation} \label{eq:pion_vff_polynomial}
    F_\pi^V(s) = (1+\alpha_V s) \, \Omega(s)
\end{equation}
with the slope parameter $\alpha_V=0.118(2)\GeV^{-2}$, using the polynomial freedom of the Omnès representation.\footnote{While this significantly improves upon the lineshape of the pion vector form factor in the vicinity of the $\rho$ resonance peak, compared to just an Omnès function $F_\pi^V(s) = \Omega(s)$, it messes up the high-energy behavior, which is supposed to be $F_\pi^V(s) \asymp s^{-1}$ up to logarithmic corrections~\cite{Lepage:1979zb,Farrar:1979aw,Donoghue:1996bt}. However, as we will later cut off the form factor dispersion relation before this becomes relevant, this does not harm us.} The slope parameter is determined by fitting to $|F_\pi^V(s)|^2$ data at timelike $s$ from measurements of $\tau^- \to \pi^- \pi^0 \nu$ decays in the CLEO and Belle experiments~\cite{CLEO:1999dln,Belle:2008xpe} as well as at spacelike $s$ obtained from $\pi H$ scattering in the NA7 experiment~\cite{NA7:1986vav}.\footnote{The spacelike data from the NA7 experiment~\cite{NA7:1986vav} also contain isospin-breaking isoscalar contributions, which we assume to be small, however, and neglect for our purposes.} In order to avoid the impact of inelastic effects on our fit, we cut off the fit range at $s \leq 1\GeV^2$. The fit has a $\chi^2/{\rm d.o.f.}=3.9$, arising from the fact that the modified IAM employed produces a slightly too narrow $\rho$ resonance~\cite{Holz:2024diw}, in turn causing the peak to slightly overshoot in the fit by $\lesssim 2.5\,\%$, as visible in Fig.~\ref{fig:pionVFF_slopeparameter_fit}. Nevertheless, this constitutes a significant improvement compared to a plain Omnès function. 

\subsection{Fixing the Contact Terms} \label{sec:contact_fixing}
In the framework established so far, there is still one free parameter that we need to fix from external input, namely the prefactor $c_4$ of the two NLO magnetic contact terms $P_{M^* \bar{M}}(s)$ and $P_2(s)$ in Eqs.~\eqref{eq:TFF_contact_term} and~\eqref{eq:VFF_contact_terms}. In order to do so, we make use of the following connection of the (isovector) magnetic form factors evaluated at $s=0$ to the magnetic moments:
\begin{align} \label{eq:FF_sum_rule_magnetic_moment}
    F_{M^* \bar{M}}(0) = \mu_\text{IV}^{M} \,,\quad\quad\quad F_2(0) = \mV\cdot\mu_\text{IV}^{M^*}\,,
\end{align}
with the isovector combination $\mu_\text{IV}\equiv (\mu_+ - \mu_0)/2$. As described in the following, we can determine $\mu_\text{IV}^{D}$ from data and demand that $F_{D^* \bar{D}}(0)$ saturates this value, which, due to the linearity of the dispersion relations, uniquely determines the parameter $c_4$. 

To obtain values for the magnetic moments $\mu_+^{D}$ and $\mu_0^{D}$, we use the matrix element for the process $D^*_a \to D_a \gamma^*$,
\begin{align}
    \mathcal{M}\bigl(D^*_a&(p,\lambda) \to D_a(p^\prime) \, \gamma(q,\lambda^\prime)\bigr) \notag\\
    &= \frac{e \mu_a^D}{M_{D^*}} \epsilon^{\mu\nu\alpha\beta} \epsilon_\mu^*(\vec{q},\lambda^\prime)\, q_\nu\, \epsilon_\alpha(\vec{p},\lambda)\, p_\beta\,,
\end{align}
which leads to a width of
\begin{equation} \label{eq:width_DstarDgamma}
    \Gamma\bigl(D^*_a\rightarrow D_a\gamma\bigr) = \frac{\alpha}{3}|\mu_a|^2|\textbf{q}|^3\,. 
\end{equation}
Using the measured widths and branching fractions of the charged and neutral $D^*$ mesons~\cite{ParticleDataGroup:2024cfk}, one can extract the magnetic moment $\mu$. For the charged decay $D^{*+}\rightarrow D^+\gamma$ this results in
\begin{align}\label{eq:result_muplus}
    \abs{\mu_+^D}&= \sqrt{\frac{3\cdot\BR(D^{*+}\rightarrow D^+\gamma)\cdot\Gamma^{D^*}_{\text{tot}}}{\alpha \cdot |\textbf{q}|^3}} \notag\\
    &= 0.47(6)\GeV^{-1}\,.
\end{align}
For the uncharged decay the difficulty arises that an exact value for the total decay width of the $D^{*0}$ is not available. However, it is possible to estimate the $D^{*0}\rightarrow D^0\pi^0$ decay width theoretically. Using the Lagrangian from Eq.~\eqref{eq:HQET_Lagrangian} with the coupling $g$ from Eq.~\eqref{eq:g}, one can write
\begin{align}
    \Gamma(D^{*0}\rightarrow D^0 \pi^0)_{\mathrm{theo}}&= \frac{1}{24\pi}\frac{g^2}{\Fpi^2}|\textbf{p}_\pi|^3\,.
\end{align}
This results in the value for the magnetic moment
\begin{align} \label{eq:result_muzero}
    \abs{\mu^D_0} &= \sqrt{\frac{3\cdot\BR(D^{*0}\rightarrow D^0\gamma) \cdot\Gamma(D^{*0}\rightarrow D^0 \pi^0)_{\mathrm{theo}}}{\alpha \cdot |\textbf{q}|^3 \cdot \BR(D^{*0}\rightarrow D^0 \pi^0)}} \notag\\
    &= 1.77(4)\GeV^{-1}\,.
\end{align}
Combining the results from Eqs.~\eqref{eq:result_muplus} and~\eqref{eq:result_muzero} and choosing the same signs as in Ref.~\cite{Amundson:1992yp} leads to 
\begin{align} \label{eq:magneticmoment}
    \mu^D_\text{IV}=\frac{\mu^D_+-\mu^D_0}{2}=1.12(4)\GeV^{-1} \,.
\end{align}

Employing heavy-quark symmetry, we can also compute an estimate for the $B$ mesons' magnetic moments up to one-loop order~\cite{Amundson:1992yp}, which is done in~\ref{app:magnetic_moments}. For the isovector combination these corrections cancel exactly and yield the same value as for the $D$ mesons, 
\begin{align}
    \mu^B_\text{IV}=\frac{\mu^B_+-\mu^B_0}{2}=1.12(4)\GeV^{-1} \,.
\end{align}
As we will see in the following section, this expectation holds within the error bounds.

\subsection{Resulting Form Factors} \label{sec:FF_results}

With all parameters fixed, we can compute the form factors for $M^{(*)}\bar{M}^{(*)} \to \gamma^*$ numerically from the dispersion relation in Eq.~\eqref{eq:FF_dispersion_relation}, both for $M=D$ and $M=B$. To that end, we first introduce an upper cutoff $s_\text{cutoff}=2.0\GeV^2$ to the form factor dispersion integrals as well $s_\text{cutoff,MO}=2.2\GeV^2$ to the Muskhelishvili--Omnès integrals.\footnote{The latter cutoff is chosen slightly higher than the former as we want to avoid its cutoff effects to influence the form factor dispersion integrals.} This is reasonable as our parametrization of the $\pi\pi$ phase shift does not include the effects of higher resonances and, thus, the contributions from above this cutoff are not very meaningful. 
Information from the higher-energy region is suppressed by the integral kernel $1/[s^\prime(s^\prime-s)]$.
The specific value for the cutoff is chosen in order to optimize the saturation of the sum rule for $F_i(0)$ in Eq.~\eqref{eq:FF_sum_rule},
\begin{equation} \label{eq:FF_sum_rule_saturation}
    F_i(0) = \frac12 \bigl( F_i^+(0) - F_i^0(0) \bigr) \overset{!}{=} \frac12 ( 1 - 0 ) = \frac12\,,
\end{equation}
which is demanded by gauge invariance for the two electric form factors $F_{M\bar{M}}$ and $F_{1,M^*\bar{M}^*}$.\footnote{As for the magnetic $D^*$ form factor, anomalous effects introduce an imaginary part to $F_{1,D^*\bar{D}^*}(0)$. In contrast to the magnetic or higher multipole moments, here this effect is unphysical: the charge, as the form factor normalization with the $D^*$ resonance evaluated at its pole position on the second Riemann sheet in the complex mass plane, must not be renormalized and therefore, in particular, remains real~\cite{Gegelia:2010nmt,Bauer:2012at,Djukanovic:2013mka}. Given the tininess of the effect, we refrain from analytically continuing in the $D^*$ mass, and ignore the imaginary part in the following.} In particular, we find
\begin{align}
\begin{alignedat}{2}
    F_{D\bar{D}}(0) &= 0.421\,,& F_{B\bar{B}}(0) &= 0.400\,,\\
    F_{1,D^*\bar{D}^*}(0) &= 0.394-0.002\ii\,,\quad& F_{1,B^*\bar{B}^*}(0) &= 0.393\,,
\end{alignedat} \label{eq:NFF_sum_rule_saturation}
\end{align}
which means we have a saturation of $\Order(80\%)$.

For the NLO magnetic contact term, we obtain
\begin{align} \label{eq:c4}
    c_4 = 0.241(7)\GeV^{-2}\,,
\end{align}
via the procedure outlined in Sect.~\ref{sec:contact_fixing}, where the error is propagated from the coupling $g$ [\cf~Eq.~\eqref{eq:g}] and the magnetic moment [\cf~Eq.~\eqref{eq:magneticmoment}]. As a result, we find the following values for the magnetic moments:
\begin{align}
    \mu_\text{IV}^{D^*} &= F_{2,D^* \bar{D}^*}(0) / M_{D^*} = (1.058-0.011\ii)\GeV^{-1}\,, \notag\\
    \mu_\text{IV}^{B} &= F_{B^* \bar{B}}(0) = 1.105\GeV^{-1}\,, \notag\\
    \mu_\text{IV}^{B^*} &= F_{2,B^* \bar{B}^*}(0) / M_{B^*} = 1.094\GeV^{-1}\,.
\end{align}
As expected, these values respect heavy-quark symmetries up to a few percent, where the biggest symmetry-breaking effect stems from the anomalous effects in the $D^*$ form factors. The latter further introduce a small imaginary part.

\begin{figure*}
    \centering
    \begin{subfigure}{.48\linewidth}
        \centering
        \includegraphics[width=\linewidth]{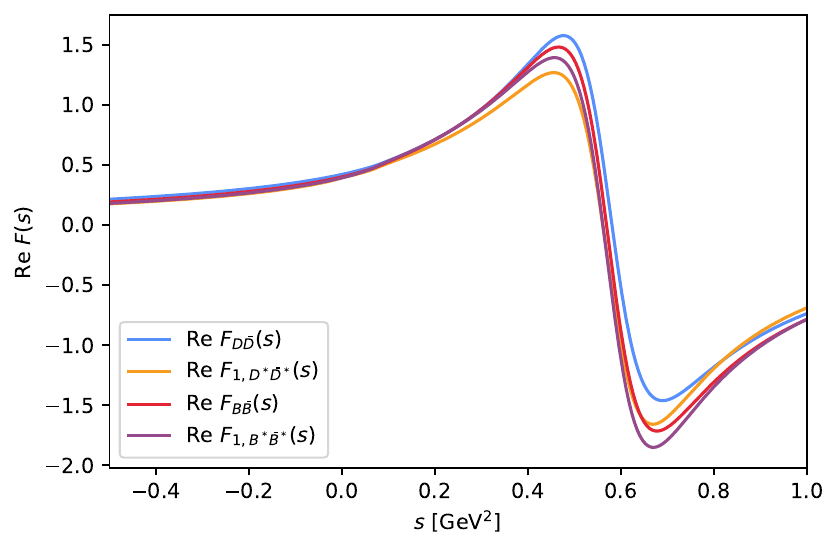}
    \end{subfigure} \hfill
    \begin{subfigure}{.48\linewidth}
        \centering
        \includegraphics[width=\linewidth]{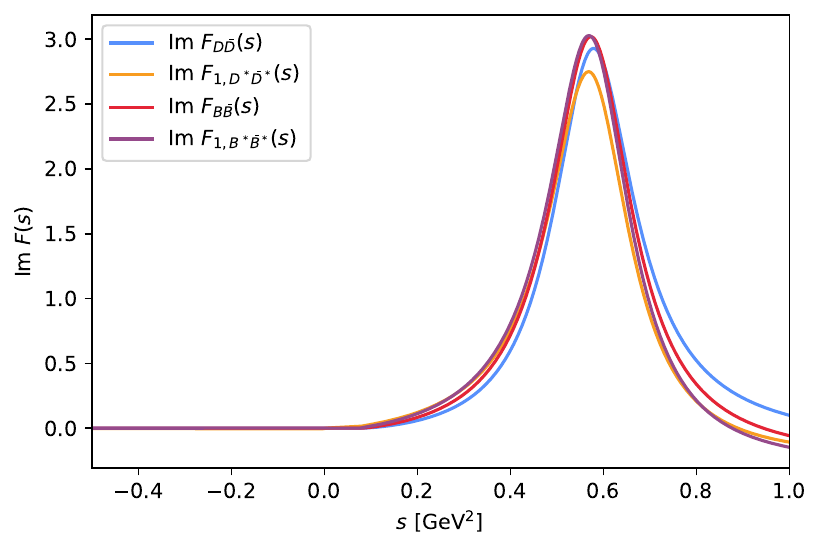}
    \end{subfigure}
    \begin{subfigure}{.48\linewidth}
        \centering
        \includegraphics[width=\linewidth]{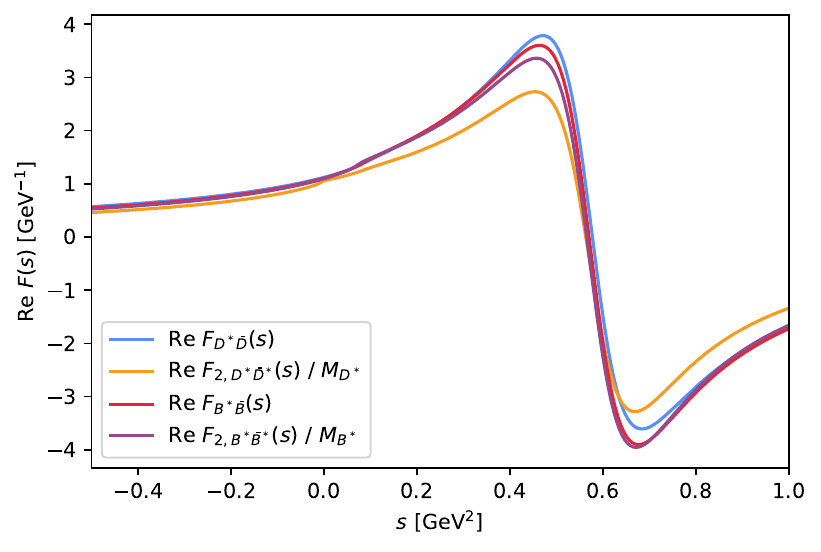}
    \end{subfigure} \hfill
    \begin{subfigure}{.48\linewidth}
        \centering
        \includegraphics[width=\linewidth]{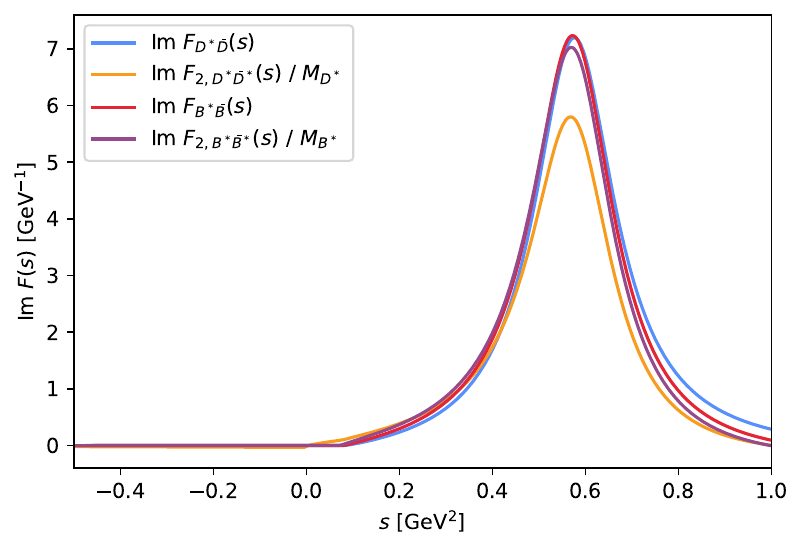}
    \end{subfigure} 
    \begin{subfigure}{.48\linewidth}
        \centering
        \includegraphics[width=\linewidth]{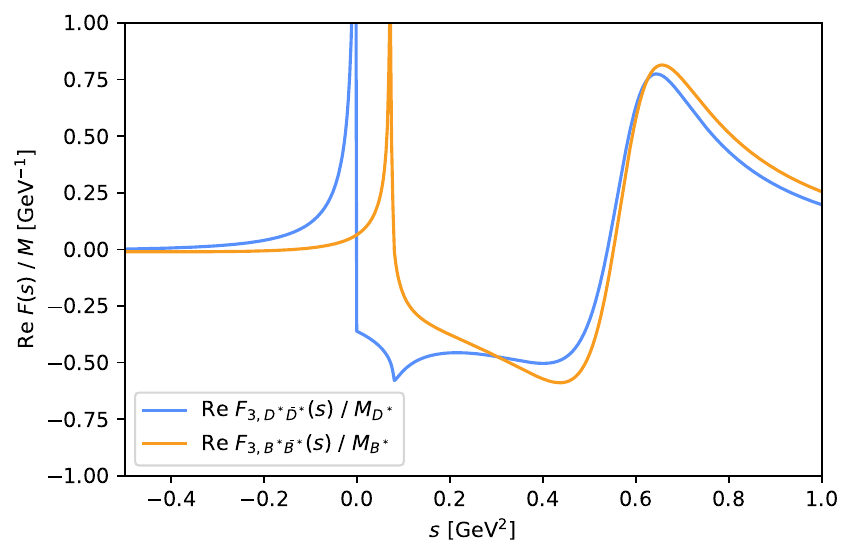}
    \end{subfigure} \hfill
    \begin{subfigure}{.48\linewidth}
        \centering
        \includegraphics[width=\linewidth]{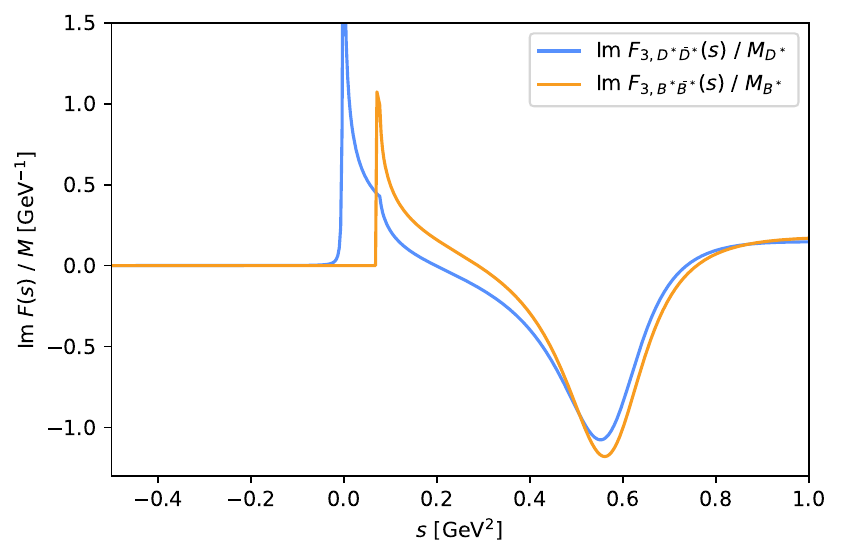}
    \end{subfigure}
    \caption{Real (left) and imaginary parts (right) of the $D^{(*)}\bar{D}^{(*)}$ and $B^{(*)}\bar{B}^{(*)}$ form factors.}
    \label{fig:MMbar_FFs}
\end{figure*}

The resulting form factors are displayed in Fig.~\ref{fig:MMbar_FFs}. Their statistical errors, propagated from the uncertainties in the input on the coupling $g$, the pion decay constant $\Fpi$, the magnetic moment $\mu_\mathrm{IV}$, as well as the IAM input parameters and the slope parameter $\alpha_V$ (\cf\ Sect.~\ref{sec:phase_input}), are not visible as they are $\lesssim 1\%$ and therefore much smaller than the systematic errors discussed in the next section. All form factors share the common features of the $\pi\pi$ threshold $s_{\text{thr}} \simeq 0.078\GeV^2$ and the $\rho$ resonance peak at $s \simeq 0.6\GeV^2$. Furthermore, for all $M^*\bar{M}^*$ form factors an anomalous threshold $s_+^{VP}$ is present below $s_{\text{thr}}$, as discussed in Sect.~\ref{sec:anomalous_thresholds}. For $M=D$ it is located close to $s=0$ at $s_+^{VP}\simeq-0.02\GeV^2$, due to the instability of $D^*\to D\pi$. The latter also leads to an imaginary part along the entire real axis, as well as the lifting of the Schwarz reflection principle due to the infinitesimal imaginary part $\mV^2\mapsto\mV^2+\ii\delta$. For $M=B$ we instead have $s_+^{VP}\simeq 0.070\GeV^2$ close to $s_{\text{thr}}$, and since $B^*\to B\pi$ is kinematically forbidden, no such imaginary part arises. In particular, this also implies that for the $D^*$ the electric and magnetic multipole moments, determined by the form factors at $s=0$, develop an imaginary part, which is particularly pronounced for the electric quadrupole moment. Similar behavior was also observed for the electromagnetic polarizabilities of $D^*$ mesons~\cite{Dang:2026bxe} as well as for higher multipole moments~\cite{Jiang:2009jn,Ledwig:2010ya,Ledwig:2011cx} and transition radii~\cite{Jiang:2009jn,Ledwig:2011cx,Junker:2019vvy,Aung:2024qmf,An:2024pip} of baryonic resonances.

One feature of these anomalous thresholds that can be immediately noticed is that for both $F_1$ and $F_2$ they merely lead to a cusp, whereas for $F_3$ they cause a logarithmic divergence. The reason for this can be understood most easily from the anomalous discontinuities in Eq.~\eqref{eq:born_disc_anom}. All of these scale with $\sim (z_P^2-1)$ and thus vanish at $s_+^{VP}$,\footnote{Recall that $s_+^{VP}$ corresponds to one of the logarithmic singularities of $\log\Bigl(\frac{1-z_P}{1+z_P}\Bigr)$, \ie,\ $z_P=\pm1$.} except for the term $\sim z_P$, which is finite and leads to a logarithmic endpoint singularity of the anomalous dispersion integral. As a consequence, the anomalous effects contribute on the order of a few percent to the $D^*$ form factors $F_1$ and $F_2$, and even less for the corresponding $B^*$ form factors, but make up a rather high percentage of the $F_3$ form factors, where they even dominate completely near the thresholds.

The importance of these anomalous effects becomes especially evident due to their implications for heavy-quark symmetry. In the heavy-quark limit we expect spin--flavor symmetry to hold. This means that, up to corrections in powers of the Compton wavelength of the heavy quarks, there should be a symmetry between the corresponding $D$ and $B$ form factors and, as it was discussed in Sect.~\ref{sec:heavy_quark_symmetry}, also between the electric-type form factors $F_{M\bar{M}}(s)$ and $F_{1,M^*\bar{M}^*}(s)$ as well as the magnetic-type form factors $F_{M^*\bar{M}}(s)$ and $F_{2,M^*\bar{M}^*}(s)$. In Fig.~\ref{fig:MMbar_FFs} one observes that this symmetry is indeed approximately fulfilled. As expected, due to the much higher mass $m_b$ compared to $m_c$, the deviations from that symmetry are much bigger for $M=D$ than for $M=B$. However, the most significant deviations occur for the $D^*$ form factors. These are caused by the anomalous effects, which do not only scale linearly with the Compton wavelengths, but are further enhanced in the proximity of the anomalous threshold. Due to the logarithmic divergence, this is most pronounced in the comparison of the two $F_{3,M^*\bar{M}^*}(s)$ form factors, but is also well visible close to threshold of the magnetic $D^*$ form factor $F_{2,D^*\bar{D}^*}(s)$.

\begin{figure*}
    \centering
    \begin{subfigure}{.48\linewidth}
        \centering
        \includegraphics[width=\linewidth]{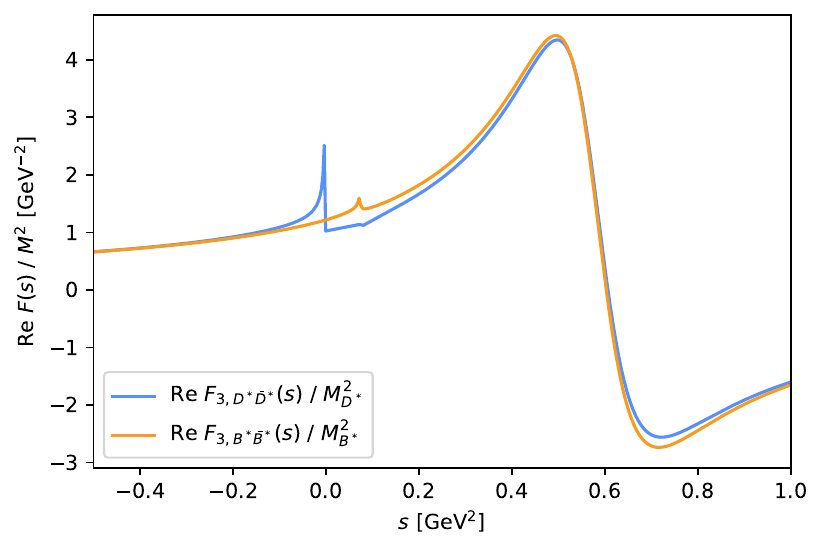}
    \end{subfigure} \hfill
    \begin{subfigure}{.48\linewidth}
        \centering
        \includegraphics[width=\linewidth]{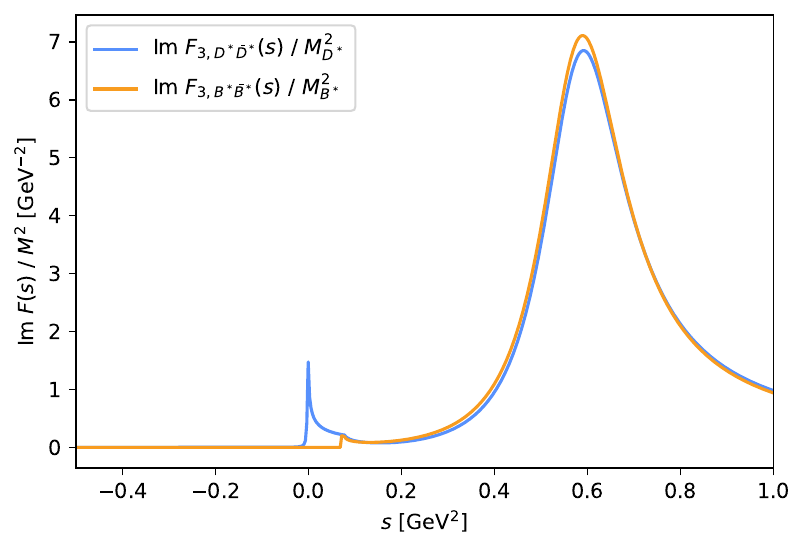}
    \end{subfigure}
    \begin{subfigure}{.48\linewidth}
        \centering
        \includegraphics[width=\linewidth]{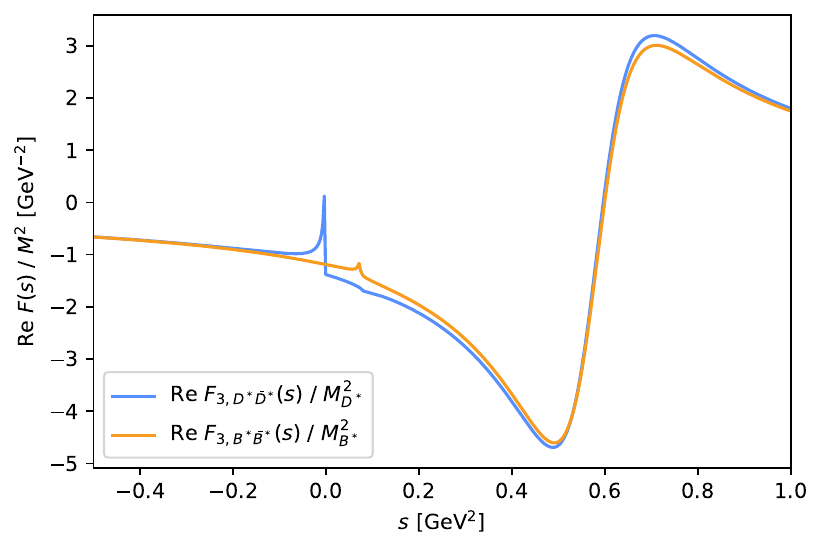}
    \end{subfigure} \hfill
    \begin{subfigure}{.48\linewidth}
        \centering
        \includegraphics[width=\linewidth]{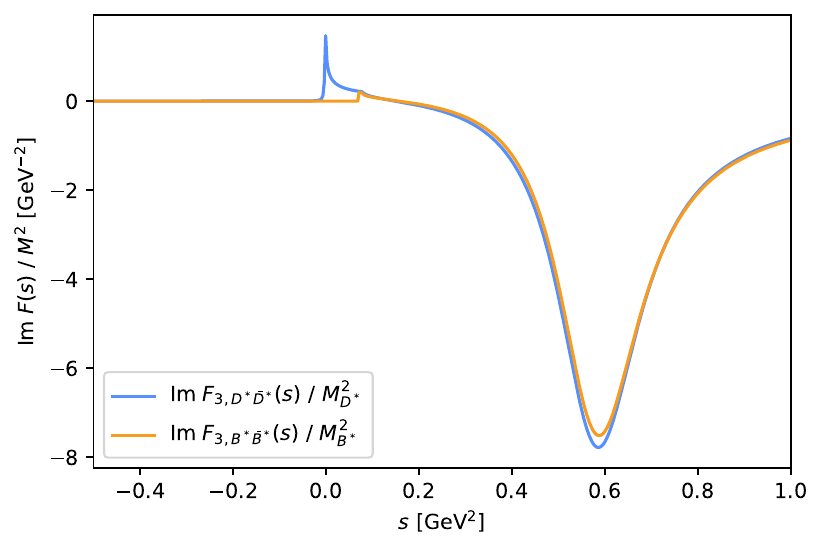}
    \end{subfigure}
    \caption{Form factor $F_3$ with an additional NNLO contact term. Upper panels: positive contact term; lower panels: negative contact term.}
    \label{fig:FF3_with_contact}
\end{figure*}

An important aspect that needs to be further addressed is the following. As discussed above, the form factor $F_3$ develops a proper logarithmic divergence at the anomalous threshold. While this lies in an unphysical region in the $B^* \bar{B}^*$ case, it occurs for small negative $q^2 \equiv -Q^2$ for $D^* \bar{D}^*$, and therefore in the physical crossed-channel $eD^*\to eD^*$ scattering region. The corresponding spin- and polarization-averaged squared scattering amplitude is given by
\begin{align}
    \overline{|\mathcal{M}|^2} &= \frac{1}{3 \mV^2 Q^2 \sin^2 \frac{\theta}{2}} \biggl[ 4\mV^4 \cos^2 \frac{\theta}{2} \Bigl( |H_{11}|^2 + 2 |H_{00}|^2 \Bigr) \notag\\
    &+ Q^2 \biggl( 2\mV^2 \cos^2 \!\frac{\theta}{2} + \bigl(Q^2 + 4\mV^2\bigr) \sin^2 \!\frac{\theta}{2} \biggr) |H_{10}|^2\biggr] ,
\end{align}
with the helicity components from Eq.~\eqref{eq:vectormeson_FF_helicity_basis} and with $\theta$ the electron refraction angle in the lab frame, $\cos\theta = \vec{k}_1 \cdot \vec{k_2} / |\vec{k}_1| |\vec{k_2}|$. Therefore, the form factor $F_3$, entering via the $H_{00}$ component, is suppressed by a factor of $Q^2$ close to zero, but it nevertheless diverges. This can be traced back to the fact that we treat $D^*$ as a stable external particle, even though the decay $D^* \to D\pi$ is kinematically allowed. When treated as a complex kinematic variable a branch cut opens up in $\mV^2$ for $\mV^2 \geq (\mP + \mpi)^2$, and as briefly discussed in Sect.~\ref{sec:anomalous_thresholds} we choose the upper branch---in agreement with causality---and introduce an infinitesimal imaginary part $\mV^2 \mapsto \mV^2 + \ii \delta$. However, this effectively means that we only give it an infinitely narrow width. We expect that if we instead analytically continued through the branch cut onto the second Riemann sheet and towards the proper pole position $\mV^2 \mapsto (M_{D^*} - \ii \Gamma_{D^*}/2)^2$, the divergence would be smeared out and rendered finite. Nevertheless, due to the physical width $\Gamma_{D^{*+}} = 83.4(1.8)\keV$~\cite{ParticleDataGroup:2024cfk} being much smaller than the distance from the threshold $M_{D^*} - M_D - \mpi = 1.73(5)\MeV$~\cite{ParticleDataGroup:2024cfk}, this would still leave behind a narrow peak and would not change the overall shape of the form factor too much. For further discussions of how this could impact observable quantities, we refer to the literature~\cite{Peierls:1961zz,Schmid:1967ojm,Ginzburg:1995js,Melnikov:1996na,Guo:2019twa}.

\begin{figure*}
    \centering
    \begin{subfigure}{.48\linewidth}
        \centering
        \includegraphics[width=\linewidth]{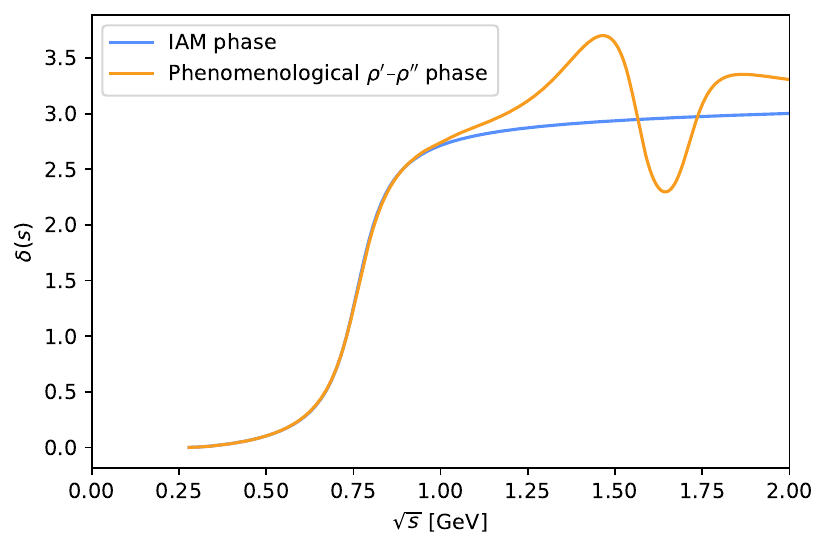}
    \end{subfigure} \hfill
    \begin{subfigure}{.48\linewidth}
        \centering
        \includegraphics[width=\linewidth]{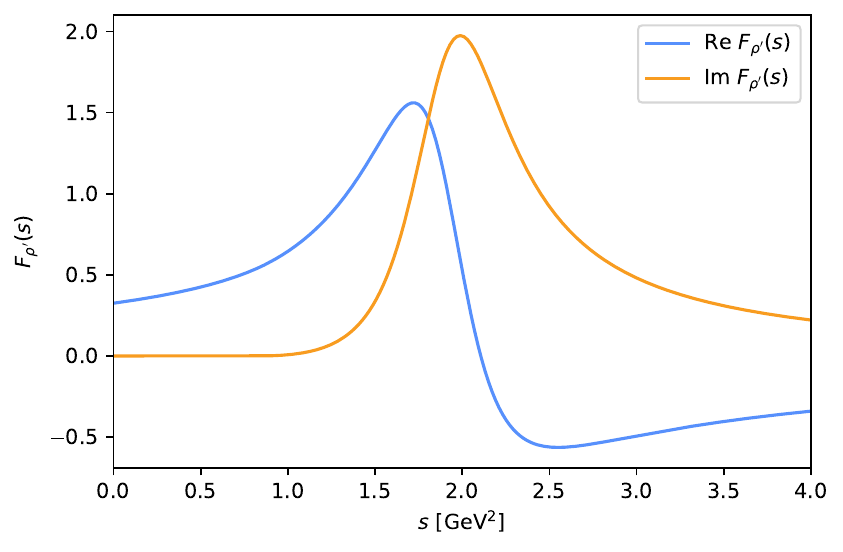}
    \end{subfigure}
    \caption{Left: Comparison between the IAM phase and the phenomenological phase from Refs.~\cite{Roig:2011iv,Schneider:2012ez} that also includes the influence of the higher $\rho^\prime$ and $\rho^{\prime\prime}$ resonances. Right: Model for the $\rho^\prime$ contribution to the form factors via the inelastic $\pi\omega$ channel, \cf\ Ref.~\cite{Zanke:2021wiq} (with arbitrary normalization).}
    \label{fig:rho_prime}
\end{figure*}

\subsection{Higher-Order Contact Terms}\label{sec:HOcontact}

In this section, we briefly discuss the scaling behavior of the $M^* \bar{M}^*$ form factors with the heavy mass $\mV$. From non-relativistic quantum mechanics---ignoring all loop effects---one would expect the natural scaling behavior for the magnetic dipole moment $\mu \sim \mV$ and the electric quadrupole moment $Q \sim \mV^2$~\cite{Kim:1973ee} as well as $\bra{M^*}j^\mu\ket{M^*} \sim \mV$ for the matrix element in total. Assuming further that in the form factor decomposition in Eq.~\eqref{eq:VV_FF_decomposition} all three terms scale in the same way and identifying $r^\mu \sim 2\mV$ and $q^\mu \sim M_\rho$, we find the following scaling behavior:
\begin{align} \label{eq:FF_scaling}
    F_1 \sim 1\,, \qquad F_2 \sim \frac{2\mV}{M_\rho}\,, \qquad F_3 \sim \frac{2 \mV^2}{M_\rho^2}\,.
\end{align}
As can be seen in Fig.~\ref{fig:MMbar_FFs}, comparing $D^{(*)}$ and $B^{(*)}$, this is nicely fulfilled for the form factors $F_1$ and $F_2$, and even the naively expected ratio at the $\rho$ peak, $|F_2/F_1| \sim 2\mV/M_\rho$, is met up to an accuracy of $\Order(5\%)$. On the other hand, this does not hold for $F_3$, whose normal contributions rather scale as $\sim \mV$ and whose anomalous contributions do not follow such a scaling behavior at all. 

While anomalous effects completely dominate near the thresholds and we, therefore, cannot expect the non-relativistic scaling to hold, they are generally suppressed in the resonance region compared to the normal contributions~\cite{Mutke:2024tww}. Had we included the NNLO contact term derived from Eq.~\eqref{NNLO-contact},
\begin{equation}
    P_3^\text{NNLO}(s) = \frac{4 d_\text{NNLO}\mV^2}{\Fpi^2} 
\end{equation}
in Eq.~\eqref{eq:VFF_contact_terms}, the expected scaling behavior of Eq.~\eqref{eq:FF_scaling} could still be fulfilled at the $\rho$ peak, where this term dominates. Contrary to the NLO contact term entering $T_2$, we would not have a way to fix the coupling strength $d_\text{NNLO}$ from data. However, purely from dimensional analysis one would expect $4d_\text{NNLO} \sim \pm 1\GeV^{-2}$ and, indeed, we find both the expected scaling behavior and the ratio at the $\rho$ peak $|F_3/F_1| \sim 2\mV^2/M_\rho^2$ to be roughly fulfilled; \cf\ Fig.~\ref{fig:FF3_with_contact}.

Note that in the threshold region, anomalous effects still provide large contributions. At $q^2 =0$, we find
\begin{align}
    F_{3,D^* \bar{D}^*}(0) &= (1.02+1.15\ii)\GeV^{-2} \mV^2 \,, \notag\\ 
    F_{3,B^* \bar{B}^*}(0) &= 1.22\GeV^{-2} \mV^2 
\end{align}
    or
\begin{align}    
    F_{3,D^* \bar{D}^*}(0) &= (-1.38+1.15\ii)\GeV^{-2} \mV^2 \,, \notag\\ 
    F_{3,B^* \bar{B}^*}(0) &= -1.19\GeV^{-2} \mV^2
\end{align}
for the positive and negative signs of the contact term, respectively.  Obviously, the imaginary parts for the $D^*$ case are by orders of magnitude different from the ones observed in the other form factors, while the real parts are dominated by the contact term.

\begin{figure*}
    \centering
    \begin{subfigure}{.447\linewidth}
        \centering
        \includegraphics[width=\linewidth]{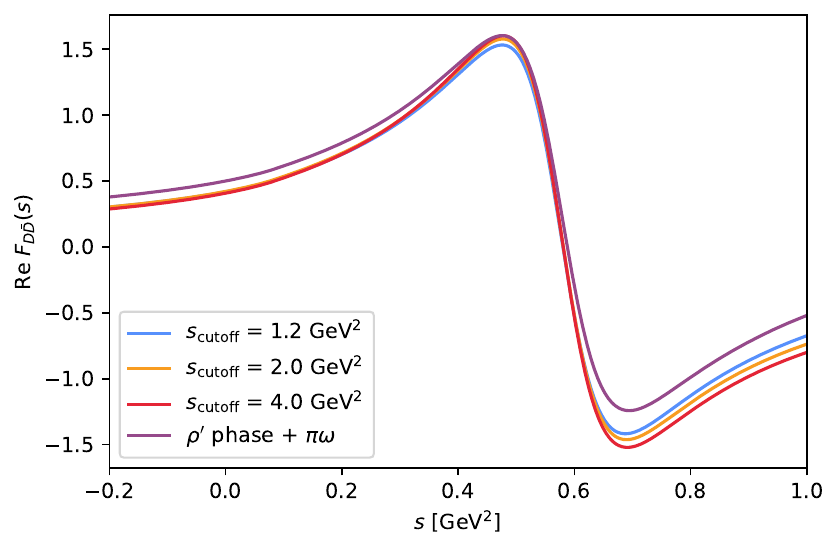}
    \end{subfigure} \hfill
    \begin{subfigure}{.447\linewidth}
        \centering
        \includegraphics[width=\linewidth]{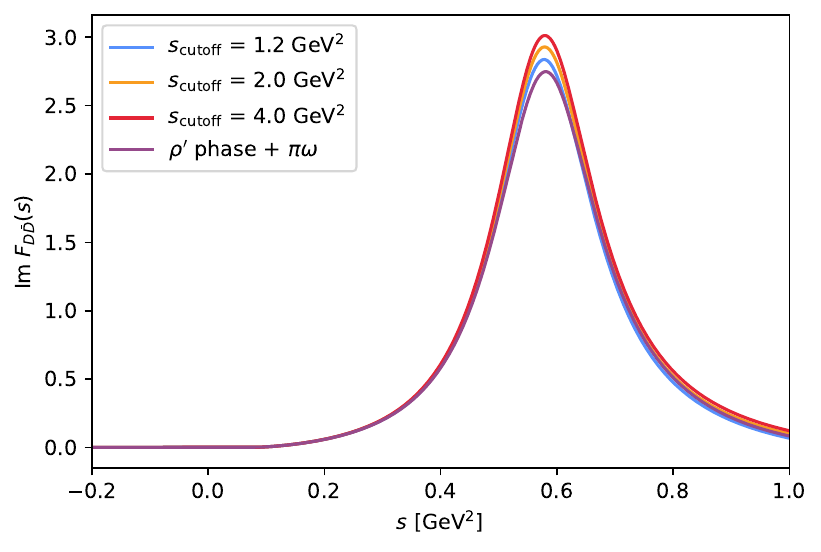}
    \end{subfigure}
    \begin{subfigure}{.447\linewidth}
        \centering
        \includegraphics[width=\linewidth]{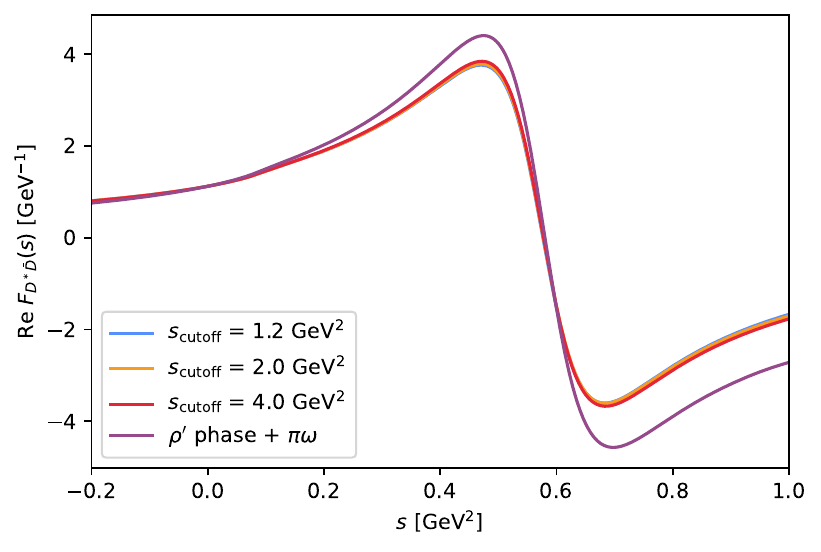}
    \end{subfigure} \hfill
    \begin{subfigure}{.447\linewidth}
        \centering
        \includegraphics[width=\linewidth]{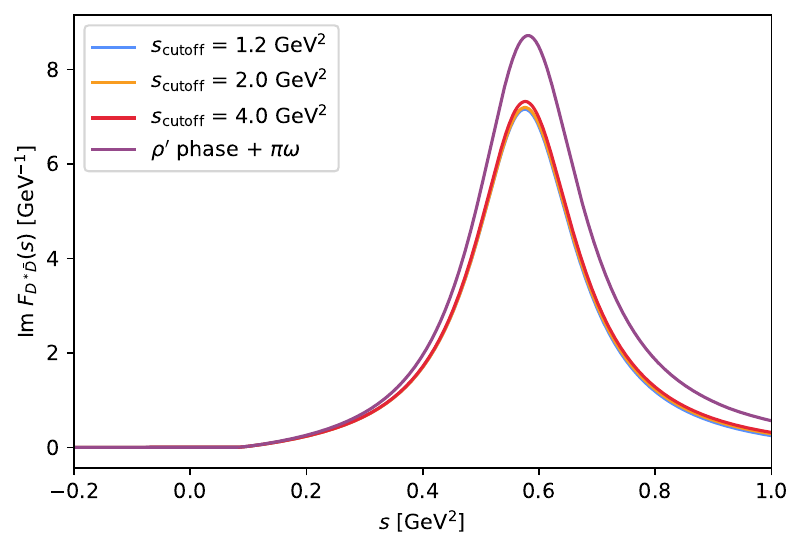}
    \end{subfigure}
    \begin{subfigure}{.447\linewidth}
        \centering
        \includegraphics[width=\linewidth]{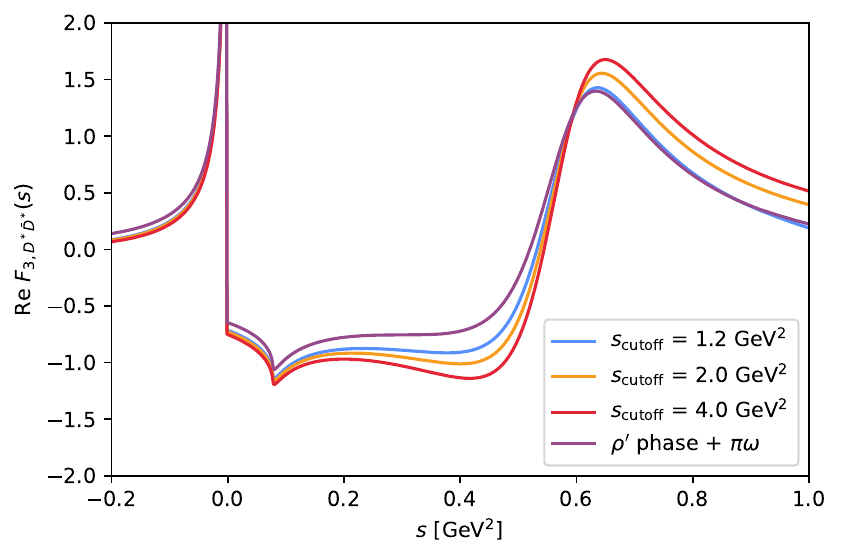}
    \end{subfigure} \hfill
    \begin{subfigure}{.447\linewidth}
        \centering
        \includegraphics[width=\linewidth]{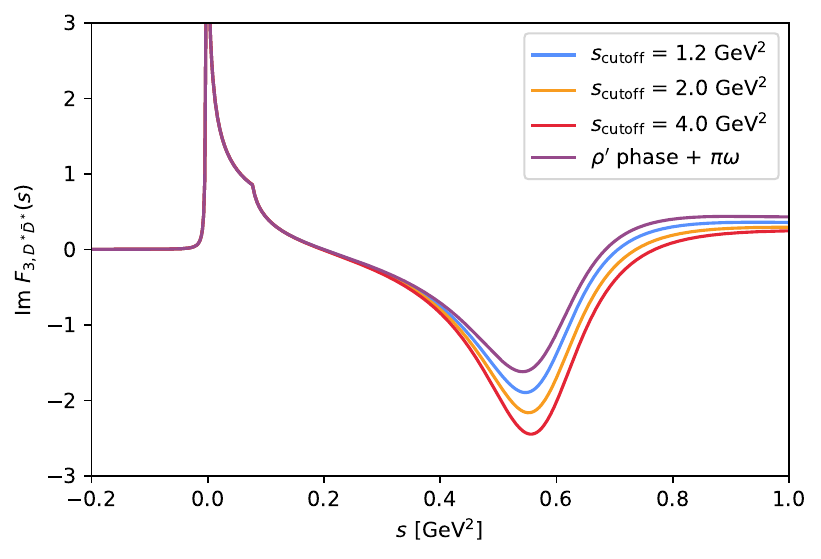}
    \end{subfigure}
    \begin{subfigure}{.447\linewidth}
        \centering
        \includegraphics[width=\linewidth]{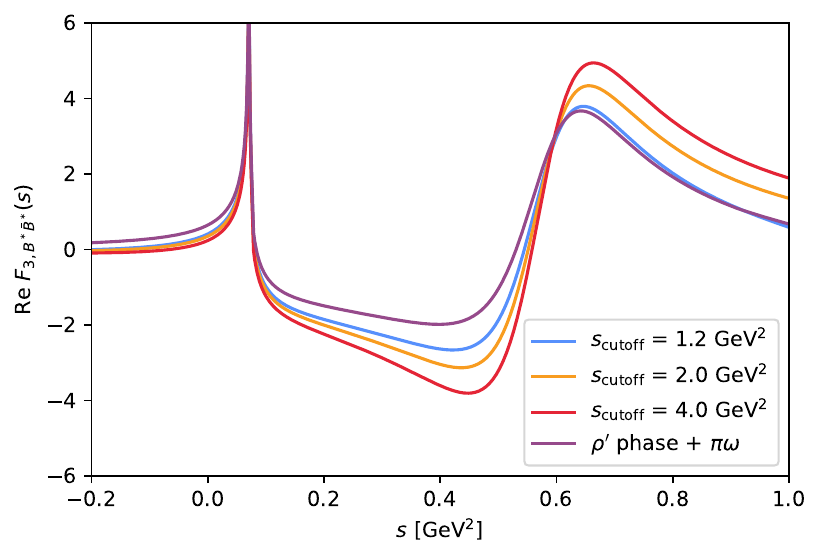}
    \end{subfigure} \hfill
    \begin{subfigure}{.447\linewidth}
        \centering
        \includegraphics[width=\linewidth]{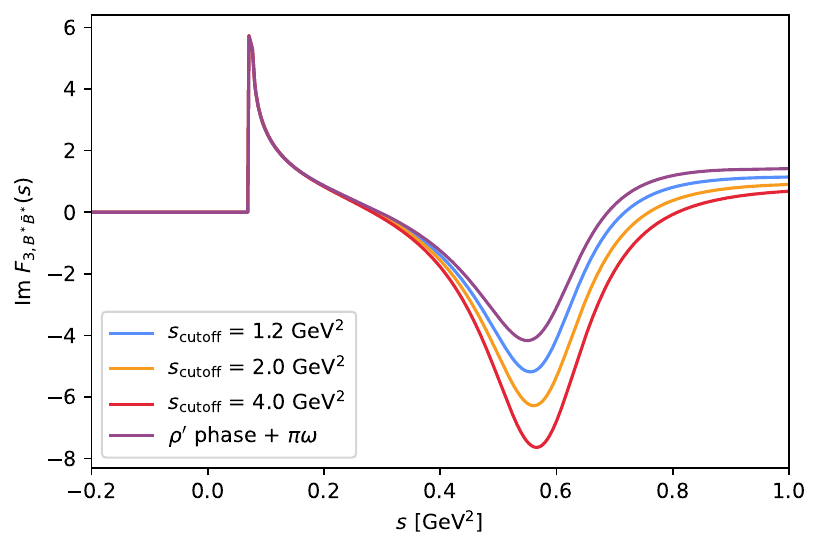}
    \end{subfigure}
    \caption{Real (left) and imaginary parts (right) of some representative $D^{(*)}\bar{D}^{(*)}$ and $B^{(*)}\bar{B}^{(*)}$ form factors for different dispersion integral cutoffs in units of $\GeV^2$. Further, for the last variant, the phase was substituted by the phenomenological phase from Refs.~\cite{Roig:2011iv,Schneider:2012ez}, including the higher $\rho^\prime$ and $\rho^{\prime\prime}$ resonances, and also an inelastic $\pi\omega$ channel was added, with a dominant $\rho^\prime$ resonance; \cf\ Ref.~\cite{Zanke:2021wiq}.}
    \label{fig:FFs_syserrs}
\end{figure*}

\subsection{Systematic Uncertainties} \label{sec:syserrs}

There are several sources of systematic uncertainties that we need to consider. First, we have chosen the cutoffs of the dispersion integrals in order to optimize the saturation of the electric charge by the sum rules in Eq.~\eqref{eq:FF_sum_rule}; however, this choice is still rather arbitrary and we have to check the effect of varying it. Second, we have only included the effects of the $\rho$ resonance in the $\pi\pi$ phase shift and disregarded higher resonances. Third, only the lightest intermediate state, namely $\pi\pi$, was considered and no inelastic effects from higher multiparticle states such as $4\pi$ have been taken into account. Finally, we have worked in the isospin limit with averaged heavy--light meson masses. However, while isospin-breaking effects would usually be quite small, the position of the $D^*$ meson's anomalous threshold close to $s=0$ is very sensitive towards small changes in the masses. In fact, the decay process $D^{*+} \to D^0 \pi^+$ is kinematically allowed while $D^{*0} \to D^+ \pi^-$ is not, meaning that the former would pick up the aforementioned imaginary part everywhere, while the latter would not. 

\begin{figure*}
    \centering
    \begin{subfigure}{0.48\textwidth}
        \centering
        \includegraphics[width=\textwidth]{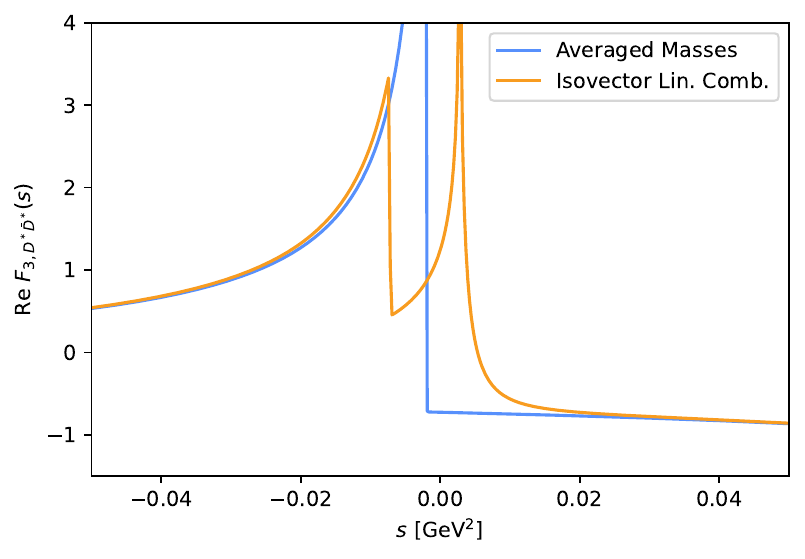}
    \end{subfigure} \hfill
    \begin{subfigure}{0.48\textwidth}
        \centering
        \includegraphics[width=\textwidth]{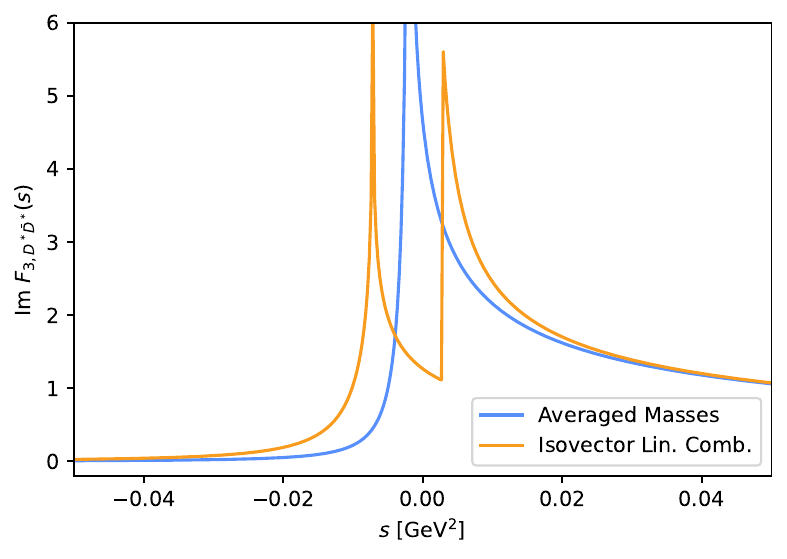}
    \end{subfigure}
    \caption{Comparison between the $D^*$ form factor $F_3$ close to the anomalous thresholds near $s=0$, both in the isospin limit and for the explicit isovector linear combination from Eq.~\eqref{eq:isospin_decomposition}, taking into account the mass difference of the charged and neutral heavy--light mesons. We show both real (left) and imaginary (right) parts.}
    \label{fig:isospin_breaking}
\end{figure*}

The second point can be addressed by replacing the $\pi\pi$ phase shift obtained via the IAM by the phenomenological form factor phase from Refs.~\cite{Roig:2011iv,Schneider:2012ez}, see Fig.~\ref{fig:rho_prime}, which also incorporates the influence of the higher $\rho^\prime$ and $\rho^{\prime\prime}$ resonances.\footnote{This prescription violates unitarity in the sense discussed in Sect.~\ref{sec:phase_input} as we cannot employ it away from the real axis, but nevertheless provides us with a reasonable error estimate.} For this replacement to be useful, we need to increase the integration cutoff to $s_\text{cutoff}=4\GeV^2$, above the two resonances, as well as set $\alpha_V=0$ in Eq.~\eqref{eq:pion_vff_polynomial}~\cite{Schneider:2012ez}. At the same time, we can address the third point by adding an inelastic $\pi\omega$ channel $\bar{F}_{\rho^\prime}(s)$ (which dominates $e^+e^-\to4\pi$ at energies up to the $\rho^\prime$ resonance~\cite{CMD-2:1998gab,Kozyrev:2019ial}),
\begin{equation} \label{eq:syserr_incoherent_sum}
    F_i^\text{(IAM phase)}(s) \mapsto F_i^\text{(pheno. phase)}(s) + c_{\rho^\prime}^{(i)} \bar{F}_{\rho^\prime}(s) \,,
\end{equation}
dominated by the $\rho^\prime$ resonance and weighted by some scaling factor $c_{\rho^\prime}^{(i)}$, to obtain a rough estimate. To model this, we employ
\begin{equation}
    F_{\rho^\prime}(s) = \int_{(\mpi+M_\omega)^2}^{\infty} \dd x \frac{\Pi_{\rho^\prime}(x)}{x-s-\ii\epsilon}\,, \quad \bar{F}_{\rho'}(s) \equiv \frac{F_{\rho'}(s)}{F_{\rho'}(0)}\,,
\end{equation}
see Fig.~\ref{fig:rho_prime}, with a spectral function for the $\rho^\prime$ resonance taken from Ref.~\cite{Zanke:2021wiq},
\begin{equation}
    \Pi_{\rho^\prime}(s) = \frac{1}{\pi} \frac{\sqrt{s} \, \Gamma_{\rho^\prime}(s)}{\bigl(s-M_{\rho^\prime}^2\bigr)^2 + s\,\Gamma_{\rho^\prime}(s)^2}\,,
\end{equation}
where the energy-dependent width is given by
\begin{align}
    &\Gamma_{\rho^\prime}(s) = \theta\bigl(s-(\mpi+M_\omega)^2\bigr)\,\frac{\gamma_{\rho^\prime\to\omega\pi}(s)}{\gamma_{\rho^\prime\to\omega\pi}\bigl(M_{\rho^\prime}^2\bigr)} \, \Gamma_{\rho^\prime}\,, \notag\\
    &\gamma_{\rho^\prime\to\omega\pi}(s) = \frac{\lambda^{3/2}\bigl(s,M_\omega^2,\mpi^2\bigr)}{s^{3/2}}\,,
\end{align}
in terms of the total width $\Gamma_{\rho^\prime}=400(60)\,\MeV$~\cite{ParticleDataGroup:2024cfk}. To determine the scaling factor $c_{\rho^\prime}^{(i)}$, we set the following demands for the sum in Eq.~\eqref{eq:syserr_incoherent_sum}. For the electric form factors we require (the real part of) the sum to saturate the electric charge; \cf\ Eq.~\eqref{eq:FF_sum_rule_saturation}. For the magnetic form factors, the magnetic moment is saturated by construction, see Sect.~\ref{sec:contact_fixing}, so we add the requirement that (the real part of) the sum should have the expected asymptotic $s^{-2}$ behavior~\cite{Farrar:1975yb,Vainshtein:1977db,Lepage:1979zb,Lepage:1980fj}. This leads to a different value for the NLO contact term,
\begin{align}
    c_4 = 0.304\GeV^{-1}\,,
\end{align}
which deviates by $\mathcal{O}(25\%)$ from the value in Eq.~\eqref{eq:c4}.  Shifts in low-order low-energy constants of such a size when including higher-order effects in matching to observables are not untypical; \cf\ the behavior of the comparable constants in pion--nucleon scattering~\cite{Hoferichter:2015tha}.
For the third $M^* \bar{M}^*$ form factors $F_{3,M^*\bar{M}^*}(s)$, we do not have a constraint on the normalization, but we use the same constraint on the asymptotic behavior as for the magnetic form factors. We obtain the numerical values for the scaling factors $c_{\rho^\prime}^{(i)}$ listed in Table~\ref{tab:couplings_rhoprime}.

\begin{table}
\renewcommand{\arraystretch}{1.5}
    \centering
    \begin{tabular}{ccc}
        \toprule
        $M$ & $D$ & $B$ \\
        \midrule
        $c_{\rho^\prime}^{(M\bar M)}$ & $0.13$ & $0.20$ \\
        $c_{\rho^\prime}^{(M^*\bar M)}$\,[GeV$^{-1}$] & $-0.28$ & $-0.23$ \\
        $c_{\rho^\prime}^{(1,M^*\bar M^*)}$ & $0.20$ & $0.23$ \\
        $c_{\rho^\prime}^{(2,M^*\bar M^*)}\,/\,\mV$\,[GeV$^{-1}$] & $-0.21$ & $-0.22$ \\
        $c_{\rho^\prime}^{(3,M^*\bar M^*)}\,/\,\mV^2$\,[GeV$^{-2}$] & $-0.015$ & $-0.009$ \\
        \bottomrule
    \end{tabular}
    \caption{Numerical values for the scaling factors $c_{\rho^\prime}^{(i)}$ of the inelastic $\rho^\prime$ contribution $\bar{F}_{\rho^\prime}(s)$ in Eq.~\eqref{eq:syserr_incoherent_sum}.}
    \label{tab:couplings_rhoprime}
\end{table}

The results for both the cutoff variation as well as the systematic error estimate via the replacement in Eq.~\eqref{eq:syserr_incoherent_sum} are shown in Fig.~\ref{fig:FFs_syserrs}. Here we have restricted ourselves to only one example for the electric and the magnetic form factors each, as the other ones display very similar behavior. As suspected, the relative systematic errors become as large as $\mathcal{O}(10\%)$ in the resonance region, dominating over the propagated statistical uncertainties, and become larger towards higher energies.

Finally, to assess the influence of working with the isospin-averaged masses, we compare our results for the $D^*$ form factors with the slightly modified version in which we explicitly use the physical masses for each charged and neutral heavy--light meson \cite{ParticleDataGroup:2024cfk},
\begin{align}
    M_{D^+} &= 1.86966(5)\GeV\,, \notag\\
    M_{D^0} &= 1.86484(5)\GeV\,, \notag\\
    M_{D^{*+}} &= 2.01026(5)\GeV\,, \notag\\
    M_{D^{*0}} &= 2.00685(5)\GeV\,,
\end{align}
before we compute the isovector combination in Eq.~\eqref{eq:isospin_decomposition}. Although this has little to no effect away from the anomalous threshold and even there is barely visible for $F_1$ and $F_2$, it is quite significant for $F_3$ in the close vicinity of $s=0$. This can be seen in Fig.~\ref{fig:isospin_breaking}. As expected, in the explicit isovector combination one anomalous threshold, corresponding to the kinematically allowed decay process $D^{*+} \to D^0 \pi^+$, sits at negative $s$ and creates an imaginary part even below the threshold, while a second one, corresponding to the kinematically forbidden process $D^{*0} \to D^+ \pi^-$, occurs for small positive $s$. 

\subsection{Radii} \label{sec:radii}

\begin{table}
\renewcommand{\arraystretch}{1.5}
    \centering
    \begin{tabular}{ccc}
        \toprule
        $M$ & $D$ & $B$ \\
        \midrule
        $\langle r^2_{M\bar M}\rangle$\,[fm$^2$] & $0.394(2)(^{7}_{8})(^{5}_{0})$ & $0.416(2)(^{\phantom{0}9}_{10})(_{3}^{0})$ \\[1mm]
        $\langle r^2_{M^*\bar M}\rangle$\,[fm$^2$] & $0.483(3)(^{6}_{1})(^{76}_{\phantom{0}0})$ & $0.513(4)(^{11}_{\phantom{0}6})(^{71}_{\phantom{0}0})$ \\[1mm]
        $\langle r^2_{1,M^*\bar M^*}\rangle$\,[fm$^2$] & $0.385(2)(8)(_{1}^{0})$ & $0.442(1)(^{10}_{\phantom{0}9})(_{6}^{0})$ \\[-1mm]
        & $~~~+\ii\, 0.267(6)(0)(0)$ & \\[1mm]
        $\langle r^2_{2,M^*\bar M^*}\rangle$\,[fm$^2$] & $0.388(2)(^{8}_{4})(^{59}_{\phantom{0}0})$ & $0.546(5)(^{13}_{\phantom{0}7})(^{67}_{\phantom{0}0})$ \\[-1mm]
        & $~+\ii\, 0.84(4)(1)(_{3}^{0})$ & \\
        \bottomrule
    \end{tabular}
    \caption{Numerical values for the mean squared radii $\langle r_i^2\rangle$ defined in Eq.~\eqref{eq:def-radii}. 
    The uncertainties quoted in brackets denote, in order:
    the errors propagated from the uncertainties in the input on the coupling $g$ and the magnetic moment $\mu_{\text{IV}}^{D}$; 
    the (asymmetric) systematic errors arising from varying the cutoff between $s_{\text{cutoff}} = 1.2\GeV^2$ and $s_{\text{cutoff}} = 4\GeV^2$;
    and the one-sided deviations obtained by substituting the input phase by the phenomenological phase from Refs.~\cite{Roig:2011iv,Schneider:2012ez}, including the higher $\rho^\prime$ and $\rho^{\prime\prime}$ resonances, and also adding an inelastic $\pi\omega$ channel dominated by the $\rho^\prime$ resonance; \cf\ Ref.~\cite{Zanke:2021wiq}.}
    \label{tab:radii}
\end{table}

The expansion of form factors at low energies, near $s=0$, allows us to define (mean squared) radii that encode information on the spatial extension corresponding to the respective probes,
\begin{equation}
F_i(s) = F_i(0) \left\{ 1 + \frac{1}{6}\langle r_i^2\rangle s + \Order\left(s^2\right) \right\} \,. \label{eq:def-radii}
\end{equation}
From Eq.~\eqref{eq:FF_dispersion_relation}, sum rules for the $\langle r_i^2\rangle$ can be gleaned immediately,
\begin{align} \label{eq:radii_sum_rule}
    \langle r_i^2\rangle &= \frac{F_i(0)^{-1}}{2\pi^2} \Bigg\{\int_{4\mpi^2}^{\infty} \dd s^\prime \frac{\pcm^3(s^\prime) \, T_i(s^\prime) \left[F_\pi^V(s^\prime)\right]^*}{s^{\prime 5/2}}\notag\\
    \Bigg(&+ \int_{0}^{1} \dd x \frac{\partial s_x}{\partial x} \frac{\pcm^3(s_x) \, \disca K_i(s_x) F_\pi^V(s_x)}{s_x^{5/2}}\Bigg)\Bigg\}.
\end{align}
The numerical values obtained herefrom are collected in Table~\ref{tab:radii}.
Note that in the electric cases, we have divided by the exact normalizations dictated by charge conservation, not by the incompletely saturated sum-rule values; \cf\ Eq.~\eqref{eq:NFF_sum_rule_saturation}. 
The isovector electric radii are roughly comparable in size to the pion radius, $\langle r_\pi^2\rangle = 0.429(4)\,\text{fm}^2$~\cite{Colangelo:2018mtw}; the isovector magnetic radii tend to be somewhat larger.  
We furthermore note that the imaginary parts of the isovector radii of the $D^*$ are very large, relatively much larger than those in the normalizations; this was similarly observed for radii of unstable baryonic resonances~\cite{Jiang:2009jn,Ledwig:2011cx,Junker:2019vvy,Aung:2024qmf,An:2024pip}.
Given that already the normalization $F_{3,M^*\bar M^*}(0)$ is a higher-order effect, with huge intrinsic uncertainties due to the unknown contact term [\cf\ Sect.~\ref{sec:HOcontact}] and even isospin breaking [\cf\ Sect.~\ref{sec:syserrs}], we refrain from showing the radii corresponding to $F_3$.

\section{Couplings to the \texorpdfstring{$\rho(770)$}{ρ(770)}}\label{sec:rho-couplings}

In this section, we want to use our knowledge about the $M^{(*)}\bar{M}^{(*)}\to\pi\pi$ $P$-wave amplitudes to extract information on the coupling of the $\rho$ resonance to the $M^{(*)}\bar{M}^{(*)}$ states. This information is encoded in its pole position and the residues on the second Riemann sheet of the amplitudes in a model-independent way.

\begin{table*}
\renewcommand{\arraystretch}{1.5}
    \centering
    \begin{tabular}{cccccc}
    \toprule
        $M$ & \multicolumn{2}{c}{$D$} & \multicolumn{2}{c}{$B$} & \cite{Casalbuoni:1996pg,Isola:2003fh,Liu:2009qhy,Lee:2009hy,Liu:2010xh,Liu:2019stu} \\
    \midrule
        Coupling & $|g|$ & $\arg g\,[^\circ]$ & $|g|$ & $\arg g\,[^\circ]$ & $|g|$ \\
    \midrule
        $\grhomm$ & $2.73(1)(^{7}_{8})$ & $2.9(3)(^{8}_{6})$ & $2.88(1)(^{11}_{12})$ & $9.1(4)(^{15}_{11})$ & 2.6 \\
        $\grhomstarm\,[\mathrm{GeV}^{-1}]$ & $1.65(6)(^{3}_{1})$ & $6.3(4)(^{3}_{4})$ & $1.68(6)(^{3}_{2})$ & $10.3(6)(6)$ & 1.64 \\
        $\grhomstarmstar$ & $2.63(2)(9)$ & $14.3(5)(^{16}_{12})$ & $2.92(1)(12)$ & $13.6(5)(^{19}_{14})$ & 2.6 \\
        $\frhomstarmstar\,[\mathrm{GeV}^{-1}]$ & $1.34(5)(^{3}_{1})$ & $14.8(8)(6)$ & $1.63(5)(^{4}_{1})$ & $13.3(7)(^{7}_{6})$ & 1.64 \\
        $\hrhomstarmstar\,[\mathrm{GeV}^{-2}]$ & $0.139(3)(^{14}_{11})$ & $-148(0)(^{6}_{4})$ & $0.059(1)(^{10}_{\phantom{0}7})$ & $-157(0)(^{6}_{5})$ & -- \\
    \midrule
        $\hrhomstarmstar\bigl(d_\mathrm{NNLO}=+0.25\GeV^{-2}\bigr)\,[\mathrm{GeV}^{-2}]$ & $1.540(1)(^{17}_{18})$ & $-10.31(3)(^{1}_{3})$ & $1.595(1)(^{\phantom{0}9}_{12})$ & $-8.09(1)(^{1}_{3})$ & -- \\
        $\hrhomstarmstar\bigl(d_\mathrm{NNLO}=-0.25\GeV^{-2}\bigr)\,[\mathrm{GeV}^{-2}]$ & $1.756(1)(^{18}_{17})$ & $175.84(3)(^{8}_{5})$ & $1.697(1)(^{11}_{\phantom{0}9})$ & $173.96(1)(^{4}_{2})$ & -- \\
    \bottomrule
    \end{tabular}
    \caption{Coupling constants of the various $M^{(*)}\bar{M}^{(*)}$ states to the $\rho$ resonance, extracted from pole parameters. The central values are determined using the cutoff $s_\text{cutoff} = 2\GeV^2$. In the first parentheses we quote the errors propagated from the uncertainties in the input on the coupling $g$ and the magnetic moment $\mu_\mathrm{IV}^D$. In the second parentheses we list the (asymmetric) systematic errors arising from varying the cutoff between $s_\text{cutoff} = 1.2\GeV^2$ and $s_\text{cutoff} = 4\GeV^2$. The couplings $\hrhomstarmstar$ are also displayed for varying input values for the higher-order contact term $d_\text{NNLO}$ as discussed in the main text. For comparison, we also list expected absolute values based on chiral and heavy-quark symmetries combined with vector-meson dominance~\cite{Casalbuoni:1996pg,Isola:2003fh,Liu:2009qhy,Lee:2009hy,Liu:2010xh,Liu:2019stu}.}
    \label{tab:couplings_rho}
\end{table*}

To establish the connection to heavy-meson effective theories, we want to parametrize these residues in terms of coupling constants that appear in an effective heavy-meson Lagrangian. Here we describe the $\rho$ as a vector-meson triplet $\rho_\mu$~\cite{Klingl:1996by},
\begin{equation}
    \rho^\mu = \begin{pmatrix}
        \rho^0 & \sqrt{2}\rho^+ \\
        \sqrt{2}\rho^- & -\rho^0
    \end{pmatrix}_\mu \,,
\end{equation}
and then can write an effective Lagrangian as
\begin{align}
	\mathcal{L} &= g_1\tr\bigl(\bar{H}_a H_b v^\mu \left(\rho_\mu\right)_{ba} \bigr) \notag\\
    &+ g_2\tr\bigl(\bar{H}_a H_b \sigma^{\mu\nu} \left(\rho_{\mu\nu}\right)_{ba}\bigr) \notag\\
    &+ \ii  g_3\tr\Bigl(
    \partial_\nu\left[\bar{H}_a \dvec \partial_{\mu} H_b\right]
    \sigma^{\mu\nu} v^\lambda \left(\rho_\lambda\right)_{ba}\Bigr) \notag\\
	&= 2g_1 \bigl( P^{*\nu} v^\mu \left(\rho_\mu\right) P_{\nu}^{*\dagger} -  P v^\mu \left(\rho_\mu\right) P^\dagger\bigr) \notag\\
	&+ 4\ii g_2  \Bigl( P^*_{\mu} \left(\rho^{\mu\nu}\right) P^{*\dagger}_\nu \notag\\
    &\qquad + \left[ P \left( \tilde{\rho}^{\mu\nu}\right) P_{\mu}^{*\dagger} -  P_{\mu}^*  \left(\tilde{\rho}^{\mu\nu}\right) P^\dagger \right] v_\nu  \Bigr) \notag\\
    &+4 g_3 \left(\partial_\nu P^{*\mu\dagger}\right) v^{\lambda} (\rho_\lambda) \left(\partial_\mu P^{*\nu}\right) \,,
    \label{eq:Lagr-rho}
\end{align}
where $\rho_{\mu\nu} = \partial_\mu \rho_{\nu} - \partial_\nu \rho_{\mu}$
and $\tilde{\rho}_{\mu\nu} =\frac{1}{2} \epsilon_{\mu\nu\alpha\beta}\rho^{\alpha\beta}$.
Using the replacement rules from Eq.~\eqref{eq:HQET_relativistic_replacement}, we obtain the following covariant effective Lagrangians:
\begin{align}
	\mathcal{L}_{\rho MM} &= \ii\grhomm \!\left( P  \rho^\mu  \partial_\mu P^\dagger - P^\dagger  \rho^\mu  \partial_\mu P \right) , \notag\\
	\mathcal{L}_{\rho M^* M} &= 4 \grhomstarm \Bigl( \partial_\mu P^*_\nu  \tilde{\rho}^{\mu\nu}  P^\dagger 
    - \partial_\mu P^{*\dagger}_\nu  \tilde{\rho}^{\mu\nu}  P \Bigr)\,, \notag\\
	\mathcal{L}_{\rho M^* M^*} &= \ii\grhomstarmstar \Bigl( P^{*\nu}  \rho^\mu  \partial_\mu P^{*\dagger}_\nu 
    - P^{*\dagger\nu}  \rho^\mu  \partial_\mu P^*_\nu \Bigr) \notag\\
	&+ 4 \ii \mV \frhomstarmstar P^*_\mu \rho^{\mu\nu} P^{*\dagger}_\nu \notag\\
    &+2\ii\hrhomstarmstar \Bigl( \partial_\lambda\partial_\nu P^{*\mu}  \rho^\lambda  \partial_\mu P^{*\dagger\nu}\notag\\
    &\hspace{55pt}- \partial_\nu P^{*\mu}  \rho^\lambda  \partial_\lambda\partial_\mu P^{*\dagger\nu} \Bigr)\,. 
    \label{eq:covariant-Lagr-rho}
\end{align}
Here, comparison between Eqs.~\eqref{eq:Lagr-rho} and \eqref{eq:covariant-Lagr-rho} shows that in the heavy-quark-symmetry limit, we have the relations
\begin{align}
\grhomm = \grhomstarmstar &= g_1 \,, \notag\\
\grhomstarm = \frhomstarmstar &= g_2 \,, \notag\\
\hrhomstarmstar &= g_3 \,.
\end{align}
We introduce the different couplings for the purpose to be able to distinguish (small) symmetry-breaking effects in the following.
In addition, the effective Lagrangian for the $\rho$ resonance coupling to pions contains the term~\cite{Klingl:1996by}
\begin{equation}
	\mathcal{L}_{\rho\pi\pi} = \grhopipi \rho^0_\mu \left(\pi^+\partial^\mu\pi^- - \pi^- \partial^\mu\pi^+\right) \,.
\end{equation}
With this, we obtain the following forms for the $T$-matrices on the second Riemann sheet close to the $\rho$ pole:
\begin{align}
	T_{M\bar{M}}^{\text{II}}(s) &= \frac{2\grhopipi\grhomm}{s_\rho-s}\,, \notag\\
	T_{M^*\bar{M}}^{\text{II}}(s) &= \frac{8\grhopipi\grhomstarm}{s_\rho-s}\,, \notag\\
	T_{1,M^*\bar{M}^*}^{\text{II}}(s) &= \frac{2\grhopipi\grhomstarmstar}{s_\rho-s}\,, \notag\\
	T_{2,M^*\bar{M}^*}^{\text{II}}(s) &= \frac{8\mV\grhopipi\frhomstarmstar}{s_\rho-s}\,, \notag\\
	T_{3,M^*\bar{M}^*}^{\text{II}}(s) &= \frac{8\mV^2\grhopipi\hrhomstarmstar}{s_\rho-s}\,.
\end{align}
Note that these also fulfill the expected scaling behavior in terms of the heavy mass $M_V$ that was discussed in Sect.~\ref{sec:HOcontact} [\cf\ Eq.~\eqref{eq:FF_scaling}].

Following Ref.~\cite{Hoferichter:2017ftn}, we can therefore extract the couplings as follows:
\begin{align}
	\grhomm &= \grhopipi \frac{\ii s_\rho \sigma^3_\pi(s_\rho)}{48\pi} T_{M\bar{M}}(s_\rho)\,, \notag\\
	\grhomstarm &= \grhopipi \frac{\ii s_\rho \sigma^3_\pi(s_\rho)}{192\pi} T_{M^*\bar{M}}(s_\rho)\,, \notag\\
	\grhomstarmstar &= \grhopipi \frac{\ii s_\rho \sigma^3_\pi(s_\rho)}{48\pi} T_{1,M^*\bar{M}^*}(s_\rho)\,, \notag\\
	\frhomstarmstar &= \grhopipi \frac{\ii s_\rho \sigma^3_\pi(s_\rho)}{192\pi \mV} T_{2,M^*\bar{M}^*}(s_\rho)\,, \notag\\
	\hrhomstarmstar &= \grhopipi \frac{\ii s_\rho \sigma^3_\pi(s_\rho)}{192\pi\mV^2} T_{3,M^*\bar{M}^*}(s_\rho)\,.
\end{align}
This has the advantage that one does not need to analytically continue the amplitudes onto the second Riemann sheet to extract the couplings, but can rather evaluate them at the pole position $s_\rho$ on the first sheet. However, in this evaluation, one has to be careful to choose the proper sheets of the logarithms appearing in the Born amplitudes $K_i(s)$; \cf\ \ref{app:analytic_structure_partial_waves}. For the $\rho$ resonance, we extract the pole parameters from the modified IAM amplitude $t_{\text{IAM}}^{\text{mod}}(s)$~\cite{Holz:2015tcg,Holz:2024diw}: the pole position $\sqrt{s_\rho} = M_\rho - \ii \Gamma_\rho/2$ is given in terms of its mass and width 
\begin{equation}
    M_\rho = 761.1\MeV\,, \qquad \Gamma_\rho = 140.3\MeV\,,
\end{equation}
while the pole residue
\begin{equation}
	\grhopipi = \frac{48\pi}{s_\rho-4M_\pi^2} \lim_{s \to s_\rho} (s_\rho-s) \, t_{\text{IAM}}^{\text{mod},\text{II}}(s) 
\end{equation}
yields the coupling to the $\pi\pi$ channel
\begin{equation}
    \abs{\grhopipi} = 5.86\,, \qquad \arg (\grhopipi) = -5.0^\circ\,,
\end{equation}
sufficiently close to the most recent phenomenological determination~\cite{Hoferichter:2023mgy}.

We obtain the values for the $\rho M^{(*)} \bar{M}^{(*)}$ couplings listed in Table~\ref{tab:couplings_rho}. For comparison, we list their expected absolute values based on chiral and heavy-quark symmetries combined with vector-meson dominance~\cite{Casalbuoni:1996pg,Isola:2003fh,Liu:2009qhy,Lee:2009hy,Liu:2010xh,Liu:2019stu}; further determinations of these couplings via various methods can be found in the literature~\cite{Li:2002pp,Bracco:2001dj,Bracco:2007sg,Wang:2007ci,Bracco:2011pg,Can:2012tx,Cui:2012wk,El-Bennich:2016bno,Ballon-Bayona:2017bwk,Kim:2019rud,Aliev:2021cjt}.  We find that our results mostly agree with these expectations up to deviations of the order of a few percent. Especially for $M=B$ the expected heavy-quark symmetry $\grhomm \approx \grhomstarmstar$ and $\grhomstarm \approx \frhomstarmstar$ is approximately fulfilled. The most significant deviations from the symmetry limit occur for $M=D$ for the two $\rho D^*D^*$ couplings, where symmetry-breaking anomalous effects play an important role (\cf\ Sect.~\ref{sec:FF_results}).

\begin{sloppypar}
A special case is given by the couplings $\hrhomstarmstar$ that are closely related to the $\rho$ contribution to the form factors $F_3$, for which we already noted in Sect.~\ref{sec:HOcontact} that it has a strong dependence on the higher-order contact term $d_\text{NNLO}$ that was only estimated by dimensional analysis.  The dependence of $\hrhomstarmstar$ on $d_\text{NNLO}$ can be quantified as follows:
\begin{align}
    \hrhodstardstar\bigl(d_\mathrm{NNLO}\bigr) &= (-0.118-0.074\ii)\GeV^{-2} \notag\\
    &+ 4\, (1.633 - 0.202\ii)\, d_\mathrm{NNLO} \,, \notag\\
    \hrhobstarbstar\bigl(d_\mathrm{NNLO}\bigr) &= (-0.054-0.023\ii)\GeV^{-2} \notag\\
    &+ 4\, (1.633 - 0.202\ii)\, d_\mathrm{NNLO} \,.
\end{align}
Table~\ref{tab:couplings_rho} also includes numbers for the $\hrhomstarmstar$ for $d_\text{NNLO} = \pm 0.25 \GeV^{-2}$ for comparison, which increases their absolute values by about one order of magnitude. In contrast to the values for $d_\text{NNLO} = 0$, they approximately fulfill the expected heavy-quark symmetry.
\end{sloppypar}

The quoted errors on the couplings in Table~\ref{tab:couplings_rho} arise as follows. First, we propagate the errors from the uncertainties in the input of the coupling $g$ [\cf~Eq.~\eqref{eq:g}] and the magnetic moment $\mu_\mathrm{IV}^D$ [\cf~Eq.~\eqref{eq:magneticmoment}], which make up the largest contribution among all input parameters. Second, as it was done in Sec.~\ref{sec:syserrs} for the form factors, we vary the high-energy cutoff between $1.2\GeV^2 \leq s_\mathrm{cutoff} \leq 4.0\GeV^2$---with the central value determined by $s_\mathrm{cutoff} = 2.0\GeV^2$---to estimate (asymmetric) systematic errors. One special case arises for the couplings $\hrhomstarmstar$, which scale proportionally with $g^2$ in the absence of any contact terms and, thus, the propagated errors picked up by the phase of the coupling are identically zero. Note that, contrary to the discussion of systematic uncertainties in Sect.~\ref{sec:syserrs}, we cannot use the phenomenological form factor phase from Refs.~\cite{Roig:2011iv,Schneider:2012ez} here to estimate the influence of the higher $\rho^\prime$ and $\rho^{\prime\prime}$ resonances, as a matching analytic expression for the pion--pion $P$-wave scattering amplitude $t_1^1(s)$ would be required to extract the coupling constants.\footnote{In the future, such an estimate could be performed via a conformal parametrization~\cite{Balz:2025auk} that provides simultaneous analytic expressions for both the pion form factor and the pion--pion partial wave. However, more work needs to be done in that direction.}

\section{Summary and Outlook}\label{sec:summary}

In this paper, we have reconstructed all isovector vector form factors in the energy range up to $1\GeV$ for ground-state pseudoscalar and vector mesons with a single charm- or bottom-quark, based on a dispersion theoretical representation that takes pion--pion intermediate states into account in a model-independent way.  The required $P$-wave amplitudes for the annihilation of heavy-meson pairs into two pions are based on heavy-quark and chiral symmetries, and are subsequently unitarized using a Muskhelishvili--Omn\`es representation.  All necessary contact terms can be fixed from the requirement that the isovector magnetic moments be reproduced; only the quadrupole moments of the heavy vector mesons are unknown and can only be estimated using dimensional analysis.  The contributions of heavier intermediate states beyond two pions are estimated using various sum rule constraints.  We have discussed the effects of anomalous thresholds in quite some detail, which are due to triangle topologies generated from Born exchanges; for the heavy-vector form factors, they move onto the first Riemann sheet (for the $D^*$ even into the physical region) and show the most pronounced effects related to the quadrupole form factors, where we observe logarithmic singularities.  Also the isovector radii that parametrize the leading energy dependence of the form factors have been predicted.  As a further application, we have extracted all couplings of the $\rho(770)$ to $D$, $D^*$, $B$, and $B^*$ mesons in a model-independent way, assessing the pole residues on the second Riemann sheet.

For final predictions of experimental observables such as Dalitz decay distributions for $M^* \to M \, e^+ e^-$ or cross sections for $M e^- \to M^{(*)} e^-$,\footnote{Note that for unstable, but long-lived states, e.g., for the pseudoscalars $M$, it is conceivable to have $M$ beams directed on target electrons, the latter bound in atoms. In the corresponding way, e.g., the electric radius of the $\Sigma^-$ has been determined \cite{SELEX:2001fbx}.} one also needs the isoscalar parts of the form factors, which are beyond the scope of the present work.
In view of the fact that the lowest-lying isoscalar vector mesons with photon quantum numbers are relatively narrow---relevant here are presumably the $\omega(782)$ and to some extent the $J/\psi$---a vector-meson-dominance approach might be reasonable; \cf\ Ref.~\cite{Stamen:2022uqh} for a comparable study of kaon form factors.

In the future, it will be important to unify the description in particular of the $D$- and $D^*$-meson form factors with that in the open-charm production region~\cite{Hanhart:2023fud,Husken:2024hmi}, not the least in view of their potential impact on our understanding of charm contributions to $B$-decay anomalies~\cite{Khodjamirian:2010vf,Khodjamirian:2012rm,Gubernari:2022hxn,Ciuchini:2022wbq,Isidori:2025dkp}.    Furthermore, for the perspective of spectroscopic applications and the construction of meson-exchange potentials for heavy mesons, an analogous study of scalar form factors is also of high interest.  Work along these lines is in progress. 

\begin{acknowledgements}
\bsp
We thank Christoph Hanhart, Ulf-G.\ Mei\ss ner, Elena Nitsche, M\'eril Reboud, and Akaki Rusetsky for useful discussions. 
Partial financial support 
by the Bonn--Cologne Graduate School of Physics and Astronomy,
by the MKW NRW under funding code NW21-024-A, 
by the German Academic Scholarship Foundation,
by the Konrad-Adenauer-Stiftung e.V. with funds from the BMFTR,
and by the Swedish Research Council (Vetenskapsr\aa det) (Grant No.\ 2019-04303) is gratefully acknowledged.
\esp
\end{acknowledgements}

\appendix

\section{Polarization Vectors} \label{app:polarization}
For the processes $M^{*}(p_1,\lambda_1)\,\bar{M}^{*}(p_2,\lambda_2)\to\gamma^*(q,\lambda)$ and $M^{*}(p_1,\lambda_1)\,\bar{M}^{*}(p_2,\lambda_2)\to\pi^+(k_1)\,\pi^-(k_2)$ with $s=q^2=(p_1+p_2)^2$, we use the following sets of polarization vectors in the center-of-mass frame:
\begin{align}
    \epsilon^\mu(\vec{q},\lambda=0)&=(0,0,0,1)\,,\notag\\
    \epsilon^\mu(\vec{q},\lambda=\pm1)&=\frac{1}{\sqrt{2}}(0,\mp 1,-i,0)\,,\notag\\
    \epsilon^\mu(\vec{p}_1,\lambda_1=0)&=\frac{1}{\mV}(\pz,0,0,\sqrt{s}/2)\,,\notag\\
    \epsilon^\mu(\vec{p}_1,\lambda_1=\pm1)&=\frac{1}{\sqrt{2}}(0,\mp 1,-i,0)\,,\notag\\
    \epsilon^\mu(\vec{p}_2,\lambda_2=0)&=\frac{1}{\mV}(-\pz,0,0,\sqrt{s}/2)\,,\notag\\
    \epsilon^\mu(\vec{p}_2,\lambda_2=\pm1)&=\frac{1}{\sqrt{2}}(0,\pm 1,-i,0)\,,
\end{align}
where 
\begin{align}
    p_1^\mu &= (\sqrt{s}/2,0,0,\pz)\,, & 
    p_2^\mu &= (\sqrt{s}/2,0,0,-\pz)\,.
\end{align}
They are obtained by starting in the particle's rest frame with $\epsilon^\mu(\vec{p}=0,\lambda=0) \equiv (0,0,0,1)$, acting (if necessary) with the usual ladder operators in the spin-1 representation $(S_j)_{kl}=-\ii\epsilon_{jkl}$, and finally boosting appropriately along the $z$-axis.

\section{Feynman Rules} \label{app:feynman_rules}
In this section, we briefly list the Feynman rules for the vertices resulting from the effective Lagrangians in Eqs.~\eqref{eq:lagrangian_L3}, \eqref{eq:lagrangian_L4}, and~\eqref{eq:NLO_Lagrangian_rel}. The LO three-point vertices are
\begin{align}
    \mathcal{M}\Bigl[ &M^{*0}(p,\lambda) \to M^+(p^\prime) \, \pi^-(k) \Bigr]\notag\\
    &= \mathcal{M}\Bigl[ M^{*+}(p,\lambda) \to M^0(p^\prime) \, \pi^+(k) \Bigr] \notag\\
    &= \frac{\sqrt{2}g}{\Fpi} \sqrt{\mV\mP} \, \epsilon_\mu(\vec{p},\lambda) \, k^\mu\,, \notag\\
    \mathcal{M}\Bigl[ &M^{*0}(p,\lambda) \to M^{*+}(p^\prime,\lambda^\prime) \, \pi^-(k) \Bigr]\notag\\
    &= \mathcal{M}\Bigl[ M^{*+}(p,\lambda) \to M^{*0}(p^\prime,\lambda^\prime) \, \pi^+(k) \Bigr] \notag\\
    &= \frac{\sqrt{2}g}{\Fpi} \epsilon^{\alpha\beta\mu\nu} \, \epsilon_\alpha(\vec{p},\lambda) \, \epsilon_\beta^*(\vec{p}^\prime,\lambda^\prime) \, p_\mu \, k_\nu\,.
\end{align}
The LO four-point contact terms are
\begin{align}
    &\mathcal{M}\left[ M^+(p_1) \, M^-(p_2) \to \pi^+(k_1) \, \pi^-(k_2) \right] \notag\\
    &= -\frac{1}{4\Fpi^2}\big(p_1-p_2\big)_\mu\big(k_1-k_2\big)^\mu 
    \notag\\
    &= -
    \mathcal{M} \left[ M^0(p_1) \, \bar{M}^0(p_2) \to \pi^+(k_1) \, \pi^-(k_2) \right] \,, \notag\\[2mm]
    &\mathcal{M} \left[ M^{*+}(p_1,\lambda_1) \, M^{*-}(p_2,\lambda_2) \to \pi^+(k_1) \, \pi^-(k_2) \right] \notag\\
    &= -\frac{1}{4\Fpi^2}\big(p_1-p_2\big)_\mu\big(k_1-k_2\big)^\mu\,\epsilon_{1,\nu}\epsilon_2^\nu 
    \notag\\
    &= -
    \mathcal{M} \left[ M^{*0}(p_1,\lambda_1) \, \bar{M}^{*0}(p_2,\lambda_2) \to \pi^+(k_1) \, \pi^-(k_2) \right] , 
\end{align}
with the short-hand notation $\epsilon_1 \equiv \epsilon(\vec{p}_1, \lambda_1)$ and $\epsilon_2 \equiv \epsilon(\vec{p}_2, \lambda_2)$. The NLO four-point contact terms are
\begin{align}
    &\mathcal{M}^{\text{cont}}_{\text{NLO}} \left[ M^{*+}(p_1, \lambda_1) \, M^{-}(p_2) \to \pi^+(k_1) \, \pi^-(k_2) \right] \notag\\
    &= \frac{4 c_4}{\Fpi^2} \epsilon_{\alpha\beta\mu\nu} \, p_1^\alpha \, \epsilon_1^\beta \, k_1^\mu \, k_2^\nu  
    \notag\\
    &=-
    \mathcal{M}^{\text{cont}}_{\text{NLO}} \left[ M^{*0}(p_1, \lambda_1) \, \bar{M}^{0}(p_2) \to \pi^+(k_1) \, \pi^-(k_2) \right] , \notag\\[2mm]
    &\mathcal{M}^{\text{cont}}_{\text{NLO}}\left[ M^{*+}(p_1, \lambda_1) \, M^{*-}(p_2, \lambda_2) \to \pi^+(k_1) \, \pi^-(k_2) \right] \notag\\
    &= \frac{4 c_4 \mV}{\Fpi^2} k_1^\mu k_2^\nu (\epsilon_{2\mu} \epsilon_{1\nu} - \epsilon_{1\mu} \epsilon_{2\nu}) 
    \notag\\
    &=-
    \mathcal{M}^{\text{cont}}_{\text{NLO}} \!\left[ M^{*0}(p_1, \!\lambda_1)  \bar{M}^{*0}(p_2, \!\lambda_2) \to \pi^+(k_1)  \pi^-(k_2) \right] \!. 
\end{align}
The NNLO four-point contact term is
\begin{align}
    &\mathcal{M}^{\text{cont}}_{\text{NNLO}}\left[ M^{*+}(p_1, \lambda_1) \, M^{*-}(p_2, \lambda_2) \to \pi^+(k_1) \, \pi^-(k_2) \right] \notag\\ 
    &= \frac{d_{\rm NNLO}}{F_\pi^2} \left(p_1^\alpha-p_2^\alpha\right)\left(p_{1,\mu}-p_{2,\mu}\right)\notag\\
    &\hspace{35pt}\times\left(k_{1,\nu}k_{2,\alpha}-k_{1,\alpha}k_{2,\nu}\right)\left(\epsilon_1^\mu\epsilon_2^\nu-\epsilon_1^\nu\epsilon_2^\mu\right)\notag\\
    &=\!-
    \mathcal{M}^{\text{cont}}_{\text{NNLO}} \!\left[ M^{*0}(p_1, \!\lambda_1)  \bar{M}^{*0}(p_2, \!\lambda_2) \!\to\! \pi^+(k_1)  \pi^-(k_2) \right] \!.
\end{align}

\begin{figure*}
    \centering
    \begin{subfigure}{0.48\linewidth}
        \centering
        \includegraphics[width=\linewidth]{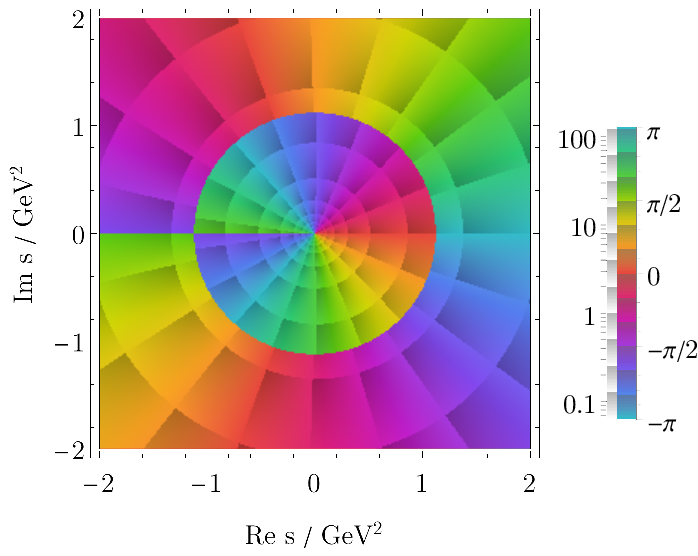}
    \end{subfigure}
    \begin{subfigure}{0.48\linewidth}
        \centering
        \includegraphics[width=\linewidth]{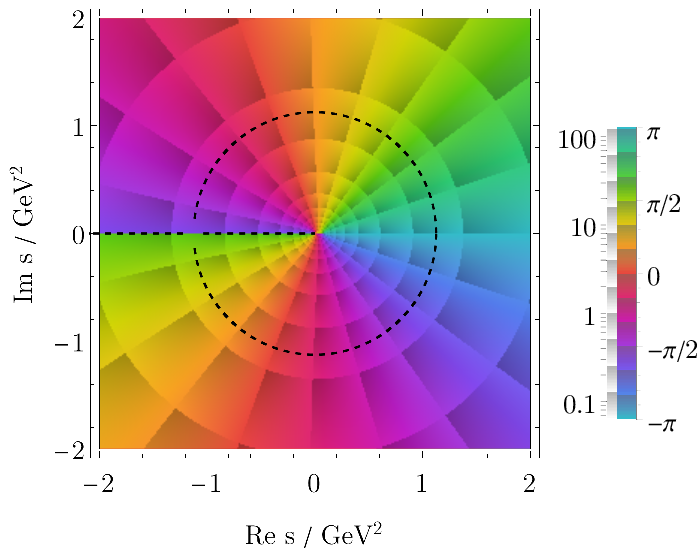}
    \end{subfigure}
    \caption{Plot of the function $g_0\bigl(s,m^2\bigr)$, defined in Eq.~\eqref{eq:partial_wave_general}, in the complex-$s$ plane for $m_1=m_2=\mV$ and $m_3=m_4=\mpi$ with $\mV=M_{D^*}$ and $m=M_D$. Different colors denote the complex phase of the function. Left: Original function. Right: Analytically continued version. The previous location of the circle-like cut as well as the left-hand cut are indicated by the dashed line.}
    \label{fig:circular_cut_DstarDstar}
\end{figure*}

\section{Analytic Structure of Partial Waves} \label{app:analytic_structure_partial_waves}
Consider a general $2\to 2$ scattering process of particles $\mathcal{P}_1(p_1) \, \mathcal{P}_2(p_2) \to \mathcal{P}_3(p_3) \, \mathcal{P}_4(p_4)$ with Mandelstam variables
\begin{equation}
    s = (p_1+p_2)^2\,, \quad t = (p_1-p_3)^2\,, \quad u = (p_1-p_4)^2\,,
\end{equation}
and masses $p_i^2=m_i^2$. Partial-wave projecting a $t$-channel (or equivalently $u$-channel) single-particle pole of the form\footnote{More generally, one could introduce a spectral density to also include $t$-channel (or $u$-channel) cuts. However, this would make the following discussion of the analytic structure more involved and is beyond our scope here. Discussions along those lines can be found in Refs.~\cite{Kennedy:1962ovz,Petersen:1969pb}.}
\begin{equation} \label{eq:LHC_pole}
    F(s,z)=\frac{a(s,z)}{t(s,z)-m^2}\,,
\end{equation}
for some analytic function $a(s,z)$, onto the $s$-channel leads to angular integrals
\begin{align}
    g_k\bigl(s,m^2\bigr) &= \frac{1}{2s\,\kappa(s)^k}\int_{-1}^{+1} \dd{z} \, \frac{P_k(z)}{t(s,z)-m^2} \notag\\
    &= \frac{-1}{\kappa(s)^{k+1}}\int_{-1}^{+1} \dd{z} \, \frac{P_k(z)}{z-Y\bigl(s,m^2\bigr)/\kappa(s)}\,,
\end{align}
with the Legendre polynomials $P_k(z)$, where we rewrote $t(s,z)-m^2 = \bigl(-Y\bigl(s,m^2\bigr)+\kappa(s)z\bigr)/2s$ with
\begin{align}
    Y\bigl(s,m^2\bigr) &= s^2 - s\bigl( m_1^2 + m_2^2 + m_3^2 + m_4^2 - 2m^2 \bigr) \notag\\
    &\hspace{10pt}+ \bigl( m_1^2 - m_2^2 \bigr) \bigl( m_3^2 - m_4^2 \bigr)\,, \notag\\
    \kappa(s) &= \lambda^{1/2}\bigl( s, m_1^2, m_2^2 \bigr) \, \lambda^{1/2}\bigl( s, m_3^2, m_4^2 \bigr)\,,
\end{align}
and included a prefactor $\sim 1/\bigl(s\,\kappa(s)^k\bigr)$ for later convenience. This integral can be evaluated in terms of the Legendre functions of the second kind $Q_k(x)$ as
\begin{equation} \label{eq:partial_wave_general}
    g_k\bigl(s,m^2\bigr) = -\frac{2}{\kappa(s)^{k+1}} \, Q_k\biggl( \frac{Y\bigl(s,m^2\bigr)}{\kappa(s)} \biggr)\,.
\end{equation}
One can express these functions as
\begin{equation}
    Q_k(x) = \frac{1}{2} P_k(x) \log\biggl(\frac{x+1}{x-1}\biggr) - W_{k-1}(x)\,
\end{equation}
with some polynomial $W_{k-1}(x)$ of degree $(k-1)$. They are analytic in the complex-$x$ plane except for a logarithmic branch cut along the real compact interval $x\in[-1,+1]$ with discontinuity
\begin{equation} \label{eq:legendre_disc}
    \disc Q_k(x) = -\ii\pi\,P_k(x)\,.
\end{equation}
Further, for $x\to\infty$ they behave as $Q_k(x) \asymp x^{-(k+1)}$.

This implies that close to the kinematic zeros of $\kappa(s)\approx 0$, we have a cancellation with the prefactors we defined, such that $g_k\bigl(s,m^2\bigr)$ behaves as $g_k\bigl(s,m^2\bigr) \sim Y\bigl(s,m^2\bigr)^{-k}$ close to these points, and therefore does not inherit any kinematic zeros or singularities from $\kappa(s)$. However, it is clear that $g_k\bigl(s,m^2\bigr)$ does inherit the logarithmic branch cuts from $Q_k(x)$ whenever $Y\bigl(s,m^2\bigr)/\kappa(s) \in [-1,+1]$, or in other words whenever there is a solution to $Y\bigl(s,m^2\bigr) = \kappa(s) z$ for some $z\in[-1,+1]$. Equivalently this means that there is a solution to $t(s,z)=m^2$, \ie, the integral kernel becomes singular. To solve this relation for $s$ at a given value of $z$, one can rewrite it into the fourth-order polynomial equation
\begin{equation}
    0 = \sum_{k=0}^{4} c_k\bigl(z^2,m^2\bigr)\,s^k\,,
\end{equation}
where the coefficients $c_k\bigl(z^2,m^2\bigr)$ are given by
\begin{align}
    c_0\bigl(z^2,m^2\bigr) &= \Delta^2 \bigl( 1-z^2 \bigr)\,, \notag\\
    c_1\bigl(z^2,m^2\bigr) &= 2\Delta \bigl( 2m^2 - R \bigr) \notag\\
    &+ 2z^2 \Bigl[ \bigl(m_1^2+m_2^2\bigr)\bigl(m_3^2-m_4^2\bigr)^2 \notag\\
    &\hspace{25pt}+ \bigl(m_3^2+m_4^2\bigr)\bigl(m_1^2-m_2^2\bigr)^2 \Bigr] \,, \notag\\
    c_2\bigl(z^2,m^2\bigr) &= 2\Delta + \bigl( 2m^2 - R \bigr)^2 \notag\\
    &-z^2 \Bigl[ \bigl(m_1^2-m_2^2\bigr)^2 + \bigl(m_3^2-m_4^2\bigr)^2 \notag\\
    &\hspace{20pt}+ 4 \bigl(m_1^2+m_2^2\bigr)\bigl(m_3^2+m_4^2\bigr) \Bigr] \,, \notag\\
    c_3\bigl(z^2,m^2\bigr) &= 2\bigl( 2m^2 - R \bigr) + 2z^2R\,, \notag\\
    c_4\bigl(z^2,m^2\bigr) &= 1-z^2\,,
\end{align}
with $\Delta=\bigl(m_1^2-m_2^2\bigr)\bigl(m_3^2-m_4^2\bigr)$ and $R=m_1^2+m_2^2+m_3^2+m_4^2$.

In the special case of interest with $m_1=m_2=\mV$ and $m_3=m_4=\mpi$, this simplifies to the quadratic equation
\begin{equation}
    0 = \sum_{k=0}^{2} \tilde{c}_k\bigl(z^2,m^2\bigr)\,s^k\,,
\end{equation}
where
\begin{align}
    \tilde{c}_0\bigl(z^2,m^2\bigr) &= 4\bigl(\mV^2+\mpi^2-m^2\bigr)^2 - 16\mV^2\mpi^2z^2\,, \notag\\
    \tilde{c}_1\bigl(z^2,m^2\bigr) &= 4m^2-4\bigl(\mV^2+\mpi^2\bigr)\bigl( 1-z^2 \bigr)\,, \notag\\
    \tilde{c}_2\bigl(z^2,m^2\bigr) &= 1-z^2\,.
\end{align}
If the mass $m$ satisfies $m^2>\mV^2-\mpi^2$, there are only real solutions for $z\in[-1,+1]$, leading to a left-hand cut $(-\infty,s_+]$ of $g_k\bigl(s,m^2\bigr)$. However, if $m^2<\mV^2-\mpi^2$, there are also complex solutions for some values of $z\in[-1,+1]$, leading instead to an additional circle-like cut\footnote{In this special case, the shape of the cut only resembles a circle, but actually is none. There are other special cases that lead exactly to a circular shape~\cite{MacDowell:1959zza}. However, there are also cases that look completely different and might even have multiple such regions.} in the complex-$s$ plane. As an example, Fig.~\ref{fig:circular_cut_DstarDstar} shows the function $g_0\bigl(s,m^2\bigr)$, as defined in Eq.~\eqref{eq:partial_wave_general}, for $\mV=M_{D^*}$ and $m=M_D$. One can analytically continue the function across this cut by using Eq.~\eqref{eq:legendre_disc}, which results in the replacement
\begin{equation}
    \log\biggl(\frac{x-1}{x+1}\biggr) \mapsto \log\biggl(\frac{x-1}{x+1}\biggr) \mp 2\pi\ii
\end{equation}
inside the circle-like region, where the negative (positive) sign needs to be used in the upper (lower) half complex-$s$ plane. The resulting analytic continuation of the previous example is shown in Fig.~\ref{fig:circular_cut_DstarDstar}. For a proper analytic continuation along the real axis, one should therefore use the replacement in Eq.~\eqref{eq:log_replacement_real}. 

\section{Magnetic Moments} \label{app:magnetic_moments}

Following the arguments of Ref.~\cite{Amundson:1992yp}, we can reproduce the $D$-meson magnetic moments as follows. We split them into contributions stemming from the photon coupling to the heavy quark $\mu^{D}_{(h)}$ and those from coupling to the light quark $\mu_{a,(l)}^{D}$,
\begin{equation} \label{eq:mu_D_from_quarks}
\mu_a^D = \mu^{D}_{(h)}+\mu_{a,(l)}^{D}\,.
\end{equation}
The former yields~\cite{Amundson:1992yp}
\begin{equation}
    \mu^{D}_{(h)} = \frac{2}{3 m_c}\left( 1-\frac{4\xi_+(1)}{m_c}\right)
\end{equation}
at second order in $1/m_c$, where the function $\xi_+$ is defined in Ref.~\cite{Luke:1990eg} and expected to be $\xi_+(1) \sim \Lambda_\text{QCD}$. The latter amounts to
\begin{align}
\mu_{+,(l)}^{D}&= + \frac{1}{3}\beta-\frac{g^2M_\pi}{8\pi \Fpi^2}\,, \notag\\
\mu_{0,(l)}^{D}&= - \frac{2}{3}\beta+\frac{g^2M_\pi}{8\pi \Fpi^2}+\frac{g^2M_K}{8\pi F_K^2} 
\end{align}  
at leading one-loop order in the chiral expansion, where $F_K = 1.1934(19) \Fpi$~\cite{FlavourLatticeAveragingGroupFLAG:2024oxs} and $\beta$ is expected to be $\beta \simeq 3\GeV$~\cite{Amundson:1992yp}. By demanding that the values $\mu_+^D=0.47(6)\GeV^{-1}$ and $\mu_0^D=-1.77(4)\GeV^{-1}$ from Eqs.~\eqref{eq:result_muplus} and~\eqref{eq:result_muzero} be reproduced, we find
\begin{equation}
    \beta = 3.11(8)\GeV\,,\quad\quad \xi_+(1)=0.545(25)\GeV\,.
\end{equation}
They indeed are of the expected orders of magnitude.

Employing heavy-quark symmetry, we now use the same values for these constants to compute 
\begin{equation} \label{eq:mu_B_from_quarks}
\mu_a^B = \mu^{B}_{(h)}+\mu_{a,(l)}^{B}
\end{equation}
with
\begin{equation}
    \mu^{B}_{(h)} = \frac{1}{3 m_b}\left( 1-\frac{4\xi_+(1)}{m_b}\right)
\end{equation}
and
\begin{align}
\mu_{+,(l)}^{B}&= + \frac{2}{3}\beta-\frac{g^2M_\pi}{8\pi \Fpi^2}-\frac{g^2M_K}{8\pi F_K^2}\,, \notag\\
\mu_{0,(l)}^{B}&= - \frac{1}{3}\beta+\frac{g^2M_\pi}{8\pi \Fpi^2} \,.
\end{align}  
We find
\begin{align}
    \mu_+^B &= 1.43(5)\GeV^{-1}\,, \notag\\
    \mu_0^B &= -0.804(23)\GeV^{-1}\,.
\end{align}
%

\bibliographystyle{utphysmod}
\bibliography{refs}

\end{document}